\documentclass[twocolumn]{aastex631}

\usepackage{multirow}

\usepackage[utf8]{inputenc}
\usepackage{natbib}
\usepackage{graphicx}
\usepackage{longtable}
\usepackage{enumerate}
\usepackage{amsmath}
\usepackage{hyperref}
\usepackage{CJK}
\usepackage[title]{appendix}
\usepackage{rotating}
\usepackage{longtable}

\usepackage{amsmath}

\def\msun{\ifmmode {\rm M}_{\mathord\odot}\else $M_{\mathord\odot}$\fi}
\def\rsun{\ifmmode {\rm R}_{\mathord\odot}\else $R_{\mathord\odot}$\fi}
\def\lsun{\ifmmode {\rm L}_{\mathord\odot}\else $L_{\mathord\odot}$\fi}

\def\c18o{C$^{18}$O}
\def\h2{H$_{2}$}
\def\13co{$^{13}$CO}
\def\n2hp{$_{2}$H$^{+}$}

\def\cm2{cm$^{-2}$}
\def\cmc{cm$^{-3}$}

\newcommand{\um}{$\mu$m}

\newcommand{\kms}{km~s$^{-1}$}

\def\um{$\mu$m}

\newcommand{\CASItD}{{\sc casi-2d}}

\def\orion{{\sc orion2}}

\shorttitle{}
\shortauthors{}

\begin{document}
\begin{CJK*}{UTF8}{gbsn}

\title{Denoising Diffusion Probabilistic Models to Predict the Density of Molecular Clouds}

\author[0000-0001-6216-8931]{Duo Xu}
\affiliation{Department of Astronomy, University of Virginia, Charlottesville, VA 22904-4235, USA}

\author[0000-0002-3389-9142]{Jonathan C. Tan}
\affiliation{Department of Astronomy, University of Virginia, Charlottesville, VA 22904-4235, USA}
\affiliation{Department of Space, Earth \& Environment, Chalmers University of Technology, SE-412 96 Gothenburg, Sweden}

\author{Chia-Jung Hsu}
\affiliation{Department of Space, Earth \& Environment, Chalmers University of Technology, SE-412 96 Gothenburg, Sweden}

\author[0000-0003-0774-9375]{Ye Zhu}
\affiliation{Department of Computer Science, Princeton University, Princeton NJ 08544, USA}
\affiliation{Department of Computer Science, Illinois Institute of Technology, Chicago, IL 60616, USA}

\email{xuduo117@virginia.edu}

\begin{abstract}

We introduce the state-of-the-art deep learning Denoising Diffusion Probabilistic Model (DDPM) as a method to infer the volume or number density of giant molecular clouds (GMCs) from projected mass surface density maps. We adopt magnetohydrodynamic simulations with different global magnetic field strengths and large-scale dynamics, i.e., noncolliding and colliding GMCs. We train a diffusion model on both mass surface density maps and their corresponding mass-weighted number density maps from different viewing angles for all the simulations. We compare the diffusion model performance with a more traditional empirical two-component and three-component power-law fitting method and with a more traditional neural network machine learning approach (\CASItD). We conclude that the diffusion model achieves an order of magnitude improvement on the accuracy of predicting number density compared to that by other methods. We apply the diffusion method to some example astronomical column density maps of Taurus and the Infrared Dark Clouds (IRDCs) G28.37+0.07 and G35.39-0.33 to produce maps of their mean volume densities.  
\end{abstract}

\section{Introduction}
\label{Introduction}

{Giant molecular clouds (GMCs) are one of the most important components of the interstellar medium (ISM) in galaxies. The ISM is the material that fills the space between stars, consisting of gas and dust, as well as cosmic rays and magnetic fields \citep{1978ppim.book.....S}. The ISM plays a critical role in the life cycle of galaxies, regulating the rate of star formation \citep[e.g.,][]{2008AJ....136.2782L,2020ARA&A..58..157T}. GMCs are particularly important because they contain the majority of the dense gas in the ISM, which is required for the formation of stars \citep[e.g.,][]{1987ARA&A..25...23S,2022ARA&A..60..319S}. The physical conditions within GMCs are highly complex and dynamic, with variations in space and time of density, temperature, velocity, and magnetic field strength and direction. These conditions can lead to a wide range of phenomena, including the formation of protostars and star clusters \citep[e.g.,][]{2007ARA&A..45..565M, 2015ARA&A..53..583H,2019ARA&A..57..227K}, and the formation of complex organic molecules \citep[e.g.,][]{2009ARA&A..47..427H,2020ARA&A..58..727J}. Consequently, investigating the physical and chemical conditions of GMCs is a crucial step towards understanding the complex physical processes that occur within the Milky Way, as well as the properties and evolution of galaxies throughout the Universe. 
}

 Among all the physical quantities of GMCs, the {density ($\rho$), i.e., mass per unit volume,}\footnote{We will also use the number density of H nuclei as a metric of density. Under the assumption of an abundance of one He nucleus for every 10 H nuclei in interstellar gas, we have a mass per H nucleus of $\mu_{\rm H} = 1.4 m_{\rm H}=2.34\times 10^{-24}\:$g. Thus $n_{\rm H} = 1\:{\rm cm}^{-3}$ is equivalent to $\rho=2.34\times 10^{-24}\:{\rm g\:cm}^{-3}$.} is one of the most fundamental properties that relates to various physical quantity estimations, such as the free-fall time, {the magnetic field strength \citep{PhysRev.81.890.2,1953ApJ...118..113C,2015A&ARv..24....4B} and chemical reaction rates \citep{1982A&A...114..245T,2010SSRv..156...13W,2017ApJ...843...38G}. The free-fall time is depends on density as $t_{\rm ff}\propto \rho^{-1/2}$. Considering measurement of magnetic fields, 
 one method commonly used 
 is the Davis-Chandrasekhar-Fermi (DCF) method \citep{PhysRev.81.890.2,1953ApJ...118..113C,2015A&ARv..24....4B}.
 This estimates the plane-of-sky (POS) component of the magnetic field
using polarized thermal dust emission \citep{1998ApJ...502L..75R,2016A&A...586A.138P}. The DCF method is based on the assumption that the magnetic field in the ISM is in a state of equipartition with the turbulent kinetic energy of the gas. This means that the magnetic field strength is proportional to the square root of the gas density and the turbulent velocity dispersion of the gas. Thus, in the DCF method, a good estimation of the gas density is required to obtain an accurate estimation of the magnetic field strength. Similarly, the gas density is a crucial factor affecting the rates of astrochemical reactions \citep{2010SSRv..156...13W,2017ApJ...843...38G}. Thus, a precise estimation of the gas density within GMCs is crucial for accurate prediction of molecular abundances and a better understanding of the chemical evolution in GMCs.
} 

However, it is difficult to quantify the number density of GMCs from observations. The traditional approach of estimating the number density of GMCs is based on observations of column density and certain assumptions on the geometry of the clouds, for example, a cylindrical geometry for filamentary structures or spherical geometry for dense cores \citep{2014prpl.conf...27A}. \citet{2021MNRAS.502.2701B,2023MNRAS.519..729B} proposed an empirical power-law to convert the observed column density to the mean number density of GMCs based on the MHD simulations from \citet{2017ApJ...835..137W}, which works decently but with noticeable scatter. Another method to constrain the number density of GMCs is utilizing density ``probes'', such as cyanoacetylene \citep[HC$_3$N,][]{1982ApJ...254..116A,1983ApJ...267..163S, 2012ApJ...756...12L}. The relative intensity of different transitions of HC$_3$N is sensitive to the number density of the cloud, which makes it possible to constrain the mean number density directly by observing multiple transitions of HC$_3$N. \citet{2012ApJ...756...12L} successfully observed $J = 2-1$ and $10-9$ transitions of HC$_3$N in Taurus B213 filament and constrained the number density of $\rm H_2$ molecules, $n_{\rm H2} \sim (1.8\pm 0.7) \times 10^{4}$~\cmc. Note, $n_{\rm H} = 2 n_{\rm H2}$ under the assumption that all H is in the form of $\rm H_2$.
Unfortunately, the line ratio of HC$_3$N can only probe the number density at a relatively narrow range, between $\sim 10^{4}$ and  $10^{6}$~\cmc\, \citep{2012ApJ...756...12L}, which limits its ability to infer the number density of the full range of structures that exist in GMCs. Consequently, a novel method to infer the number density of GMCs under a variety of physical conditions with high precision is in great demand. Machine learning makes it possible to learn from both the morphology of the cloud and their column density to infer the mean number density rather than using a simple average power-law conversion.

Machine learning has gained great popularity among astronomers. {For example, convolutional neural networks (CNNs) have been successfully applied to a series of tasks, including galaxy classification \citep{2018MNRAS.476.3661D,2021MNRAS.503.1828B}, identification of structures like protostellar outflows, stellar wind-driven bubbles and Galactic cirrus filaments \citep{2020ApJ...890...64X,2020ApJ...905..172X,2022ApJ...926...19X,2023MNRAS.519.4735S} and infer physical quantities based on observations, such as protostellar outflow inclination angles, magnetic field directions, stellar masses, exoplanet masses, galactic redshifts, and galactic star-formation rates \citep{2020ApJ...894...70L,2022MNRAS.510.4473Z,2022ApJ...941...81X,2023ApJ...942...95X,2023MNRAS.tmp...66B}. Furthermore, CNNs have also been utilized to mitigate the impact of noise in astronomical observations \citep{2022MNRAS.509..990G,2023MNRAS.tmp..637B}. For instance, \citet{2022MNRAS.509..990G} employed CNNs to remove noise and
artifacts of radio interferometric images, while \citet{2023MNRAS.tmp..637B} employed CNNs to diminish the impact of noise on various observation targets and preserve the morphology of galaxies. Meanwhile, Generative Adversarial Networks (GANs) have been applied to a variety of tasks \citep{2022ApJ...941..141H,2022MNRAS.517.4054S}. \citet{2022MNRAS.517.4054S} applied GANs to generate super-resolution and de-noised images from the $XMM-Newton$ telescope. \citet{2022ApJ...941..141H} utilized GANs to effectively deblend galaxies from HST observations.} More recently, Denoising Diffusion Probabilistic Models (DDPMs) have demonstrated their proficiency and robustness in image generation \citep{pmlr-v37-sohl-dickstein15,NEURIPS2020_diffusion}, which are suitable for the prediction task in astronomy. 

In this paper, we introduce the deep learning method denoising diffusion probabilistic models to infer the number density of GMCs from column/surface density maps. We describe the diffusion model and how we generate the training set from MHD simulations in Section~\ref{Data and Method}. Here we also introduce a convolutional neural network based machine learning approach, \CASItD\ (Convolutional Approach to Structure Identification-2D), to infer the number density from column density. In Section~\ref{Results}, we evaluate our diffusion model in predicting the number density and compare with other approaches. We also apply our diffusion model to real observations in Section~\ref{Results}. We summarize our results and conclusions in Section~\ref{Conclusions}.

\section{Data and Method}
\label{Data and Method}

\subsection{Magnetohydrodynamics Simulations}
\label{Magnetohydrodynamics Simulations}

\begin{figure*}[hbt!]
\centering
\includegraphics[width=0.93\linewidth]{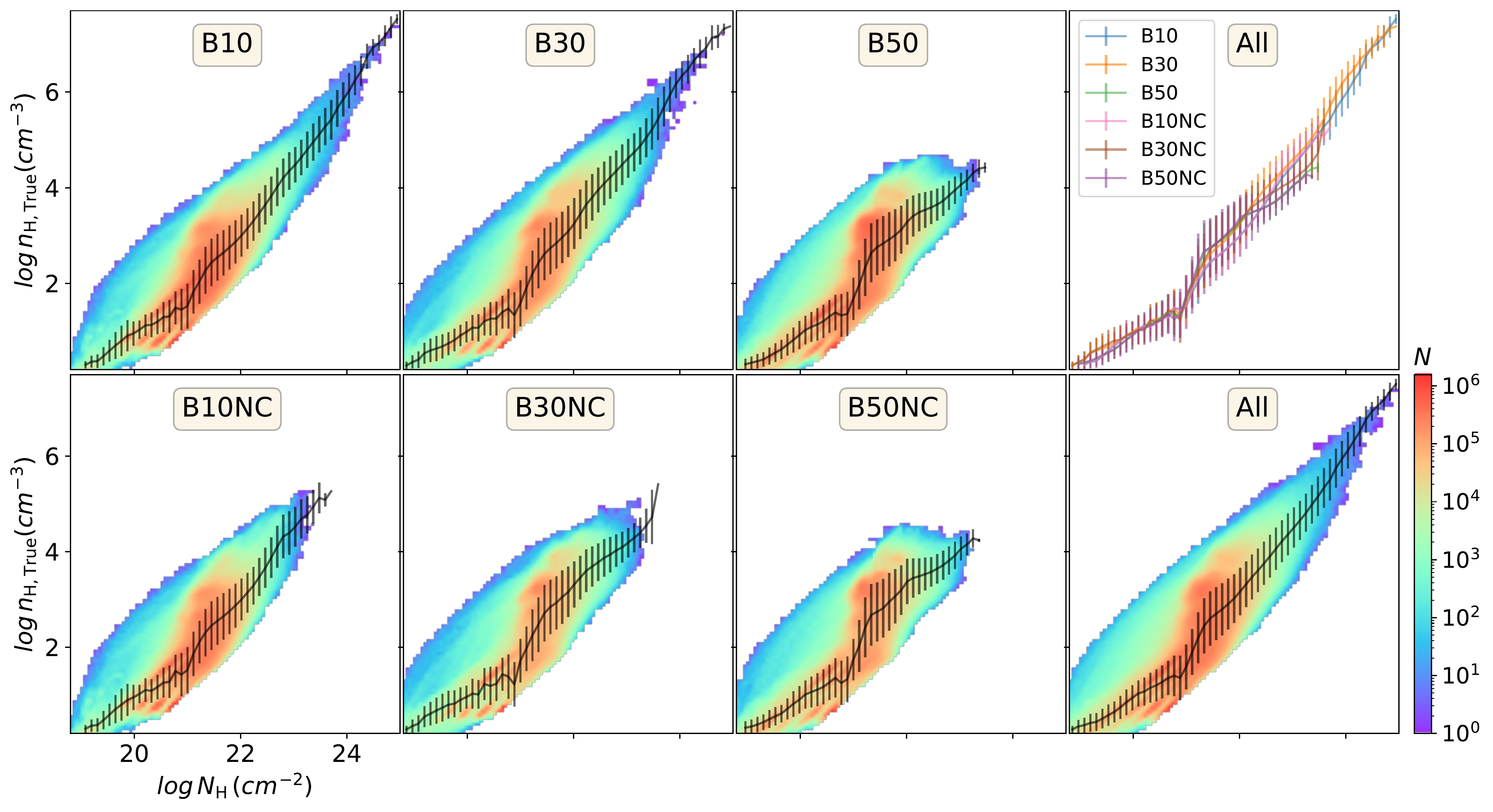}
\caption{Relation between the LOS mass-weighted number density and the column density for different simulations. The rainbow color background indicates the 2D histogram of the distribution between number density and column density for each simulation. The lines with errorbars represent the mean and their standard deviation of each column density bin.}
\label{fig.rawdata-sum-all}
\end{figure*} 

{We carry out ideal MHD simulations based on the set-up of \citet{2020ApJ...891..168W} and \citet{2023arXiv230110657H}, 
which were conducted using the MUSCL-Dedner method and HLLD Riemann solver in the adaptive mesh refinement (AMR) code Enzo \citep{2002JCoPh.175..645D, 2009ApJ...696...96W, 2014ApJS..211...19B}. The simulations include self-gravity, magnetic fields, and heating/cooling based on a photo-dissociation model, which assumes a FUV radiation field of $G_0=4$ Habings with attenuation from the n$_\textrm{H}-A_V$ relation introduced in \citet{2015ApJ...811...56W} and cosmic ray ionization rate $\zeta=10^{-16}\,{\rm s^{-1}}$. Initially, two clouds of a radius of 20~pc are initialized in a 128~pc$^3$ domain resolved by 256$^3$ cells. The clouds have initial density n$_\textrm{H}=83\:\textrm{cm}^{-3}$,  temperature $T=15\:$K, and solenoidal turbulent velocity field with $v_k^2 \propto k^{-4}, 2 \leq k \leq 20$. The gas outside the clouds has a 10 times lower density n$_\textrm{H} = 8.3\:\textrm{cm}^{-3}$ and 10 times higher temperature 150~K to balance the pressure. Note, while the GMCs have an initial temperature of 15 K, soon a multiphase temperature structure is established, e.g., with typical temperatures of $\sim 10-20\:$K at high densities ($n_{\rm H}\gtrsim 10^3\:{\rm cm}^{-3}$), $\sim 40\:$K at intermediate densities ($n_{\rm H}\sim 10^2\:{\rm cm}^{-3}$), and $\sim 1,000\:$K at low densities ($n_{\rm H}\lesssim 10\:{\rm cm}^{-3}$) \citep[see Fig. 2 of][]{2023arXiv230110657H}.

The initial magnetic field is oriented at an angle of 60$^\circ$ with respect to the collisional axis and its strength varies from 10, 30, to 50~$\mu$G in different considered cases. Four additional levels of refinement are allowed to resolve the local Jeans length with 8 cells. 
For each magnetic field, we select two different types of setup of the large-scale dynamics of the GMCs, noncolliding and colliding GMCs. In the colliding cases, the clouds have a relative velocity of 10 \kms, and are offset by 0.5$R_{\rm GMC}$. The simulations do not include star formation or feedback, so represent the structures that develop in the early phases of collapse up to the onset of star formation.  The simulations are run for 5 Myr. We take two evolutionary stages, 3 and 4 Myr, for analysis from each run. }

To enhance the diversity of the data set, we generate column density maps and their corresponding line-of-sight (LOS) mass-weighted number density across different scales by adopting different AMR levels with different physical resolutions. The image size in pixels is $128\times128$, with multiple physical scales, including 32, 16, 8, and 4 pc. In total, we have 7179 images in the data set, in which 70\% are used for the training set, and the remaining 30\% are a test set. Figure~\ref{fig.rawdata-sum-all} shows the correlation between the LOS mass-weighted number density and the column density for different simulations. Although the column density range is not the same for different simulations, it is obvious that the relation between the mass-weighted number density and column density is similar for all the different simulations.

\subsection{Machine Learning Approaches}
\label{Machine Learning Approaches}

\subsubsection{Denoising Diffusion Probabilistic Models}
\label{Denoising Diffusion Probabilistic Models}

\begin{figure*}[hbt!]
\centering
\includegraphics[width=0.93\linewidth]{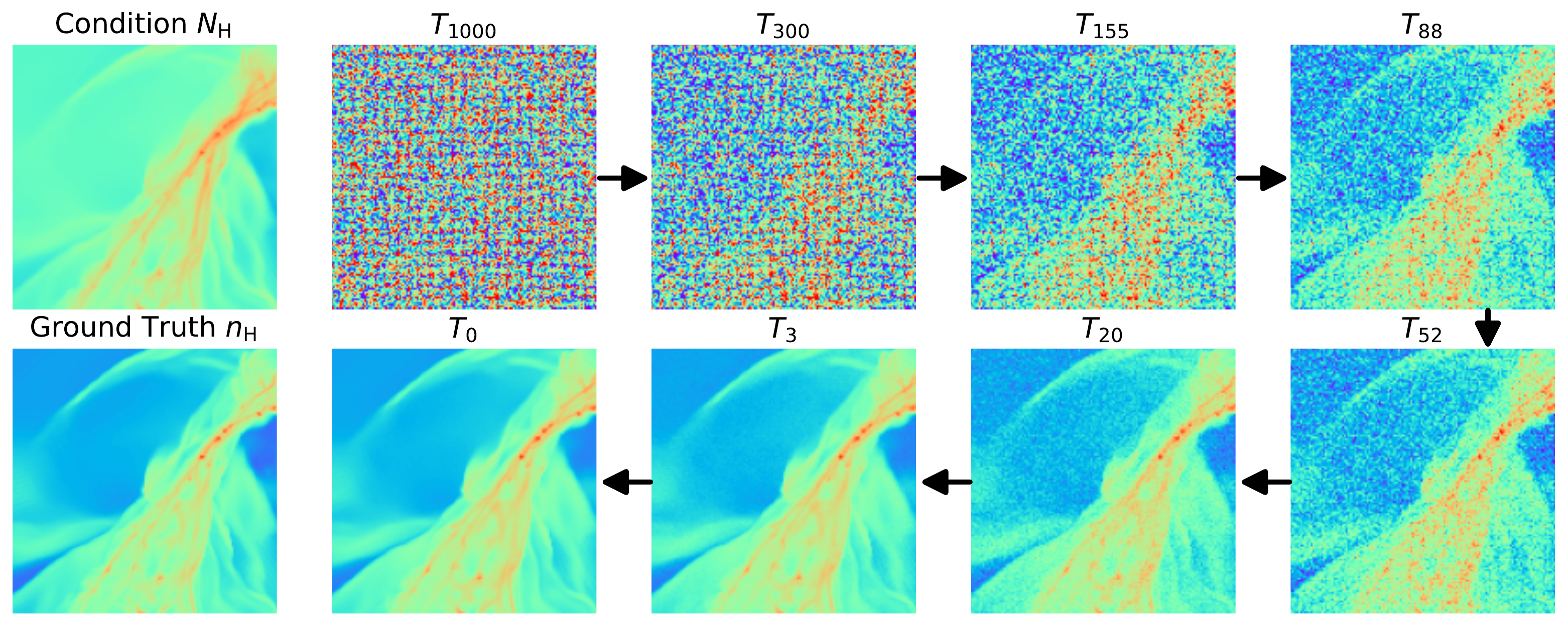}
\caption{Demonstration of the diffusion process (reverse) on a sample in the test set. }
\label{fig.diffusion-demo-example}
\end{figure*}

Denoising diffusion probabilistic models (DDPMs)~\citep{pmlr-v37-sohl-dickstein15,NEURIPS2020_diffusion}, hereafter called diffusion models for short, are the state-of-the-art generative method in deep learning and computer vision research field in recent years. 
Specifically, diffusion models have been successfully applied in synthesizing various high-fidelity data including images, video and audios~\citep{stablediff,singer2022make,zhu2022discrete,zhu2023boundary}.

The core formulation of diffusion models is inspired by non-equilibrium thermodynamics~\citep{pmlr-v37-sohl-dickstein15}, which models a stochastic Markov chain of $T$ steps in two directions.
The forward direction $q$, also known as the diffusion process, gradually adds stochastic Gaussian noises to a data sample $x_0$ following:
\begin{equation}
q(\boldsymbol{x}_{t}|\boldsymbol{x}_{t-1})=\mathcal{N}(\sqrt{1-\beta_{t}}\boldsymbol{x}_{t-1},\beta_{t}\boldsymbol{\rm I}),
\end{equation}
where $\{\beta_t\}^T_t =1$ are pre-scheduled variances.
The other direction, often referred as reverse direction or generative process $p$, denoises a noisy sample $x_T$ from a standard Gaussian distribution to a data sample $x_0$ as:
\begin{equation}
    \boldsymbol{x}_{t-1} = \frac{1}{\sqrt{1 - \beta_t}}(\boldsymbol{x}_t - \frac{\beta_t}{\sqrt{1 - \alpha_t}}\epsilon^{\theta}_t(\boldsymbol{x}_t)) + \sigma_t z_t,
\end{equation}
where $\epsilon^{\theta}_t$ is the learnable noise predictor, $z_t \sim \mathcal{N}(0, \mathbf{I})$, and the variance of the reverse process $\sigma^2_t$ is set to be $\sigma^2 = \beta$.
The actual neural network training process aims to learn the noise predictor to optimize the variational lower bound on negative log likelihood:
\begin{equation}\label{eq:vlb}
    \mathcal{L} =  \mathbb{E}_q[- \textrm{log}\: p(\boldsymbol{x}_T) - \sum_{t \geq 1} \textrm{log} \: \frac{p_{\theta}(\boldsymbol{x}_{t-1}|\boldsymbol{x}_t)}{q(\boldsymbol{x}_t | \boldsymbol{x}_{t-1})} ].
\end{equation}


There are several reasons why we adopt the diffusion model as the tool to infer the number density of GMCs. First, the diffusion models have demonstrated great potential and impressive performance in learning the data distribution in the generative field as the current mainstream method. Moreover, unlike the previous deep learning methods such as Generative Adversarial Networks (GANs)~\citep{gan}, diffusion models are well-known for the benefits of stable training, robust performance and better interpretability and traceability via the rigorous mathematical formulations as a Markov chain. Additionally, originated from the natural thermodynamics problem, the diffusion models simulate a random walk process in the data space, which shares the intrinsic alignment and consistency with most scientific problems. 


To adapt the diffusion models in our context, we deploy a diffusion model with a similar parameter setup as in~\cite{NEURIPS2020_diffusion}.
Specifically, the diffusion model has in total $T=1000$ steps, and is optimized using the variational loss in Equation~\ref{eq:vlb}, which makes valid the assumption that the reverse direction converges to the Gaussian stochastic diffusion process. In particular, we provide $\boldsymbol{x}_{c}$ as an additional input condition to make the prediction follow each individual observational sample. We train the diffusion model for 400 epochs and evaluate the performance on a sample in the test set. Figure~\ref{fig.diffusion-demo-example} shows an example of the reverse process on our test data, where Gaussian noise is gradually converted to our target after 1000 time steps.

\subsubsection{\CASItD}
\label{CASI2D}

In this section, we introduce the \CASItD\ model to predict the number density of GMCs from the column density. We adopt the same CNN architecture, \CASItD, from \citet{2019ApJ...880...83V}. \CASItD\ is an autoencoder with both residual networks \citep{he2016deep} and a “U-net” \citep{ronneberger2015u}. {\CASItD\ has two major components, the encoder part and the decoder part. The encoder part extracts the features from the input data and maps them into a lower-dimensional space, called the latent space, and then the decoder part takes this compressed representation from the latent space and attempts to reconstruct the target data. During the training process, the autoencoder is fed both input data, i.e., the column density map, and the target data, i.e., the number density map. It then learns to map the column density data to the number density data in a way that minimizes the difference between the reconstructed output and the target.} We adopt the same {training} hyperparameters as \citet{2020ApJ...890...64X,2020ApJ...905..172X}.

\section{Results}
\label{Results}

\subsection{Comparison between Different Approaches}
\label{Comparison between Different Approaches}

\begin{figure*}[hbt!]
\centering
\includegraphics[width=0.99\linewidth]{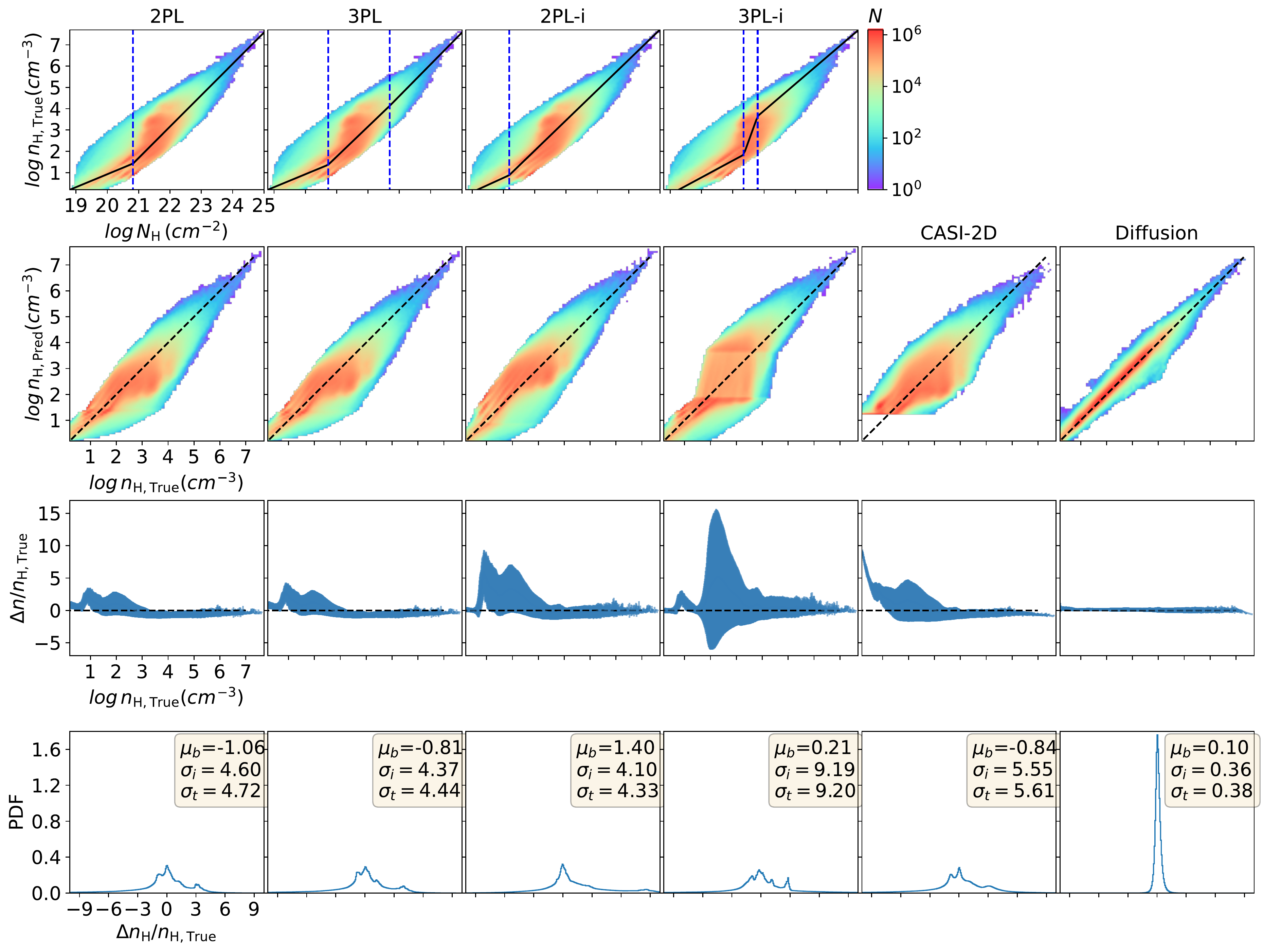}
\caption{Summary of the performance of different approaches to convert the gas column density to number density on all the data samples.}
\label{fig.diffusion-sum-all}
\end{figure*}

In this section, we adopt several different approaches to convert column density to mass-weighted number density on the LOS. {Note that when converting from $n_{\rm H}$ and $N_{\rm H}$ to a total mass per unit volume, $\rho$, or a mass per unit area (mass surface density, $\Sigma$), as mentioned above, one must also account for He (assuming $n_{\rm He}=0.1n_{\rm H}$, so that the mass per H is $\mu_{\rm H}=1.4 m_{\rm H} = 2.34\times 10^{-24}\:$g. Note, also that observational studies of molecular clouds sometimes report results for $n_{\rm H2}$, since $\rm H_2$ is the main collision partner. We will convert these estimates to $n_{\rm H}$ assuming all H is in the form of $\rm H_2$ so that $n_{\rm H}=2n_{\rm H2}$.}

We start with power-law fitting on the relation between the LOS mass-weighted number density of H nuclei ($n_{\rm H}$) and the column density of H nuclei ($N_{\rm H}$) 
for all simulations in Figure~\ref{fig.rawdata-sum-all}. 
We adopt two-components and three components power law to fit the $n_{\rm H}-N_{\rm H}$ relation, i.e., $n_{\rm H}=f(N_{\rm H})$. Meanwhile, we conduct the ``inverted'' fitting, i.e., adopting two-component and three component power laws to fit the $N_{\rm H}-n_{\rm H}$ relation $N_{\rm H}=f(n_{\rm H})$ and then derive the inverse function $n_{\rm H}=f^{-1}(N_{\rm H})$. We summarize the fitting results in Table~\ref{fitting-pl-all-sum}. We then follow the power-law fitting results to convert the column density to mass-weighted number density. We show the fitting results in Figure~\ref{fig.diffusion-sum-all}. In addition, we present the result from machine learning approaches, including \CASItD\ and the diffusion model, in Figure~\ref{fig.diffusion-sum-all}. 

Considering the results, it is obvious that there is significant dispersion between the true mass-weighted number density, $n_{\rm H,True}$, and the predicted number density, $n_{\rm H,Pred}$, that is converted by the power-law relation and that is predicted by \CASItD. The predicted number density by the diffusion model has a much smaller dispersion around the true density. We present the dispersion values between $n_{\rm H,True}$ and $n_{\rm H,Pred}$ in Figure~\ref{fig.diffusion-sum-all} for the various cases. The predicted number density by the diffusion model is an order of magnitude better than that by all the other approaches. {The 2D distribution between the actual number density and the predicted number density is shown in the second row of Figure~\ref{fig.diffusion-sum-all}. The power-law conversion approach leads to a non-uniform distribution across densities, with a systematic underestimation at low density and a systematic overestimation at moderate density. 

We proceed to measure the error distribution across the range of densities in the third row of Figure~\ref{fig.diffusion-sum-all}. It is observed that, except for the diffusion model, all other approaches have a non-uniform distribution across densities. In the last row of Figure~\ref{fig.diffusion-sum-all}, we provide an overview of the error distribution for all methods, i.e., listing the average offset ($\mu_{b}$), internal standard deviation ($\sigma_i$), and total standard deviation ($\sigma_t=\sqrt{\mu_{b}^{2}+\sigma_{i}^{2}}$). It can be seen that the diffusion model has the lowest offset bias and the smallest variation compared to the other methods. 
For the diffusion model the mean offset is at a level of 10\% and the total standard deviation, which indicates the total deviation from the true value, is at the level of 38\%.
}

\begin{figure*}[hbt!]
\centering
\includegraphics[width=0.93\linewidth]{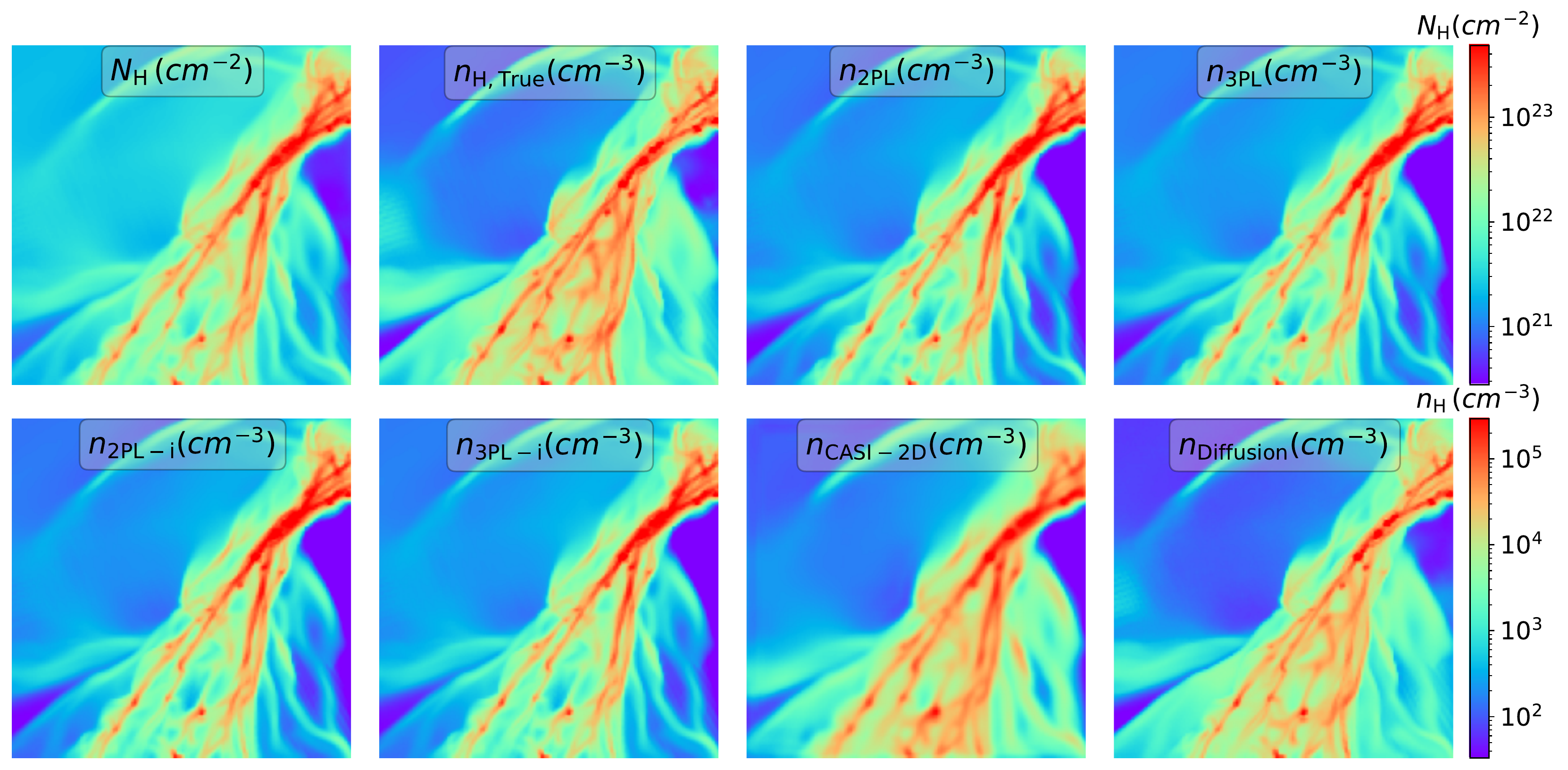}
\caption{Comparison between different approaches to converting gas column density to number density on a sample in the test set. }
\label{fig.comp-example-all}
\end{figure*}

To better visualize the performance between different approaches, we show a sample image from the test set and apply different approaches to obtain the predicted number density in Figure~\ref{fig.comp-example-all}. Although the column density map is similar to the true number density map, there is noticeable difference at relatively high column densities, where more structures are highlighted in the true number density map but appear faint in the column density map. The power-law conversion approach cannot reconstruct the structures that have relatively high number densities. \CASItD\ is able to regenerate some of the high density structures, but due to the intrinsic convolution manipulation, the number density map predicted by \CASItD\ is smoother compared to the true number density map. The number density map predicted by the diffusion model is able to recreate the actual structures across a wide density range. 

{The performance of the proposed diffusion model exhibits an improvement of one order of magnitude compared to \CASItD, which is a convolutional neural network similar to Variational Autoencoders (VAEs)~\citep{vae-kingma2013auto}.
In the context of machine learning, we discuss the reasons for the significant improvement achieved by our proposed diffusion model as follows. The DDPMs are formulated based on the Markov stochastic process and model a random walk in the data space, which aligns with most existing physics problems and is consistent with intrinsic properties of the natural world. In contrast, CNNs and VAEs were originally designed for image classification and generation tasks in computer vision and lack explicit connections to the physical world. For instance, the observed structure of GMCs is likely to be shaped, at least in part, by turbulent motions that involve compressions in a series of quasi-random directions. Then the overall mass surface density map is constructed by summing a series of quasi-independent patches of volume density along the line of sight.
Thus, inferring the raw mass-weighted number density distribution based on the observed column density inherits the basic concepts of diffusion models. Furthermore, with respect to information loss, DDPMs maintain the same data dimensionality throughout the entire denoising (i.e., prediction) process, thus better preserving the information conveyed by the original data. In contrast, CNNs and VAEs involve dimension reduction and information compression during training, resulting in inevitable information loss for the prediction objective. In terms of traceability and interpretability, we employ pre-defined Gaussian transition kernels to introduce and remove noise at each diffusion step in DDPMs. This provides us with superior traceability and interpretability for the data transition compared to CNNs and VAEs, whose traceability relies on a relatively vague gradient descent optimization direction.
}

\begin{table}[]
  
\begin{center}
  \caption{Power-Law Fitting Results$^a$ \label{fitting-pl-all-sum}}
 \begin{tabular}{cccc}
\hline
\hline
Label & Components   & Break Points & Power Indices \\
\cline{1-4} 
2PL & 2 & 20.82 &  0.61/1.47  \\ 
3PL & 3 & 20.73/22.69 &  0.61/1.41/1.49  \\
2PL-i & 2 &  20.19  &  0.67/1.42 \\
3PL-i & 3 & 21.35/21.79 &  0.79/4.06/1.26 \\
\cline{1-4} 
\multicolumn{4}{p{0.97\linewidth}}{Notes:}\\
\multicolumn{4}{p{0.97\linewidth}}{$^a$ Label, number of power-law components, break points of power-law fittings in log scale, and power indices of each power-law component.}
\end{tabular}%
\end{center}
\end{table}%

\subsection{Test on $Herschel$ Observations of Taurus B213}
\label{Test on $Herschel$ Observations of Taurus B213}

\begin{figure*}[hbt!]
\centering
\includegraphics[width=0.7\linewidth]{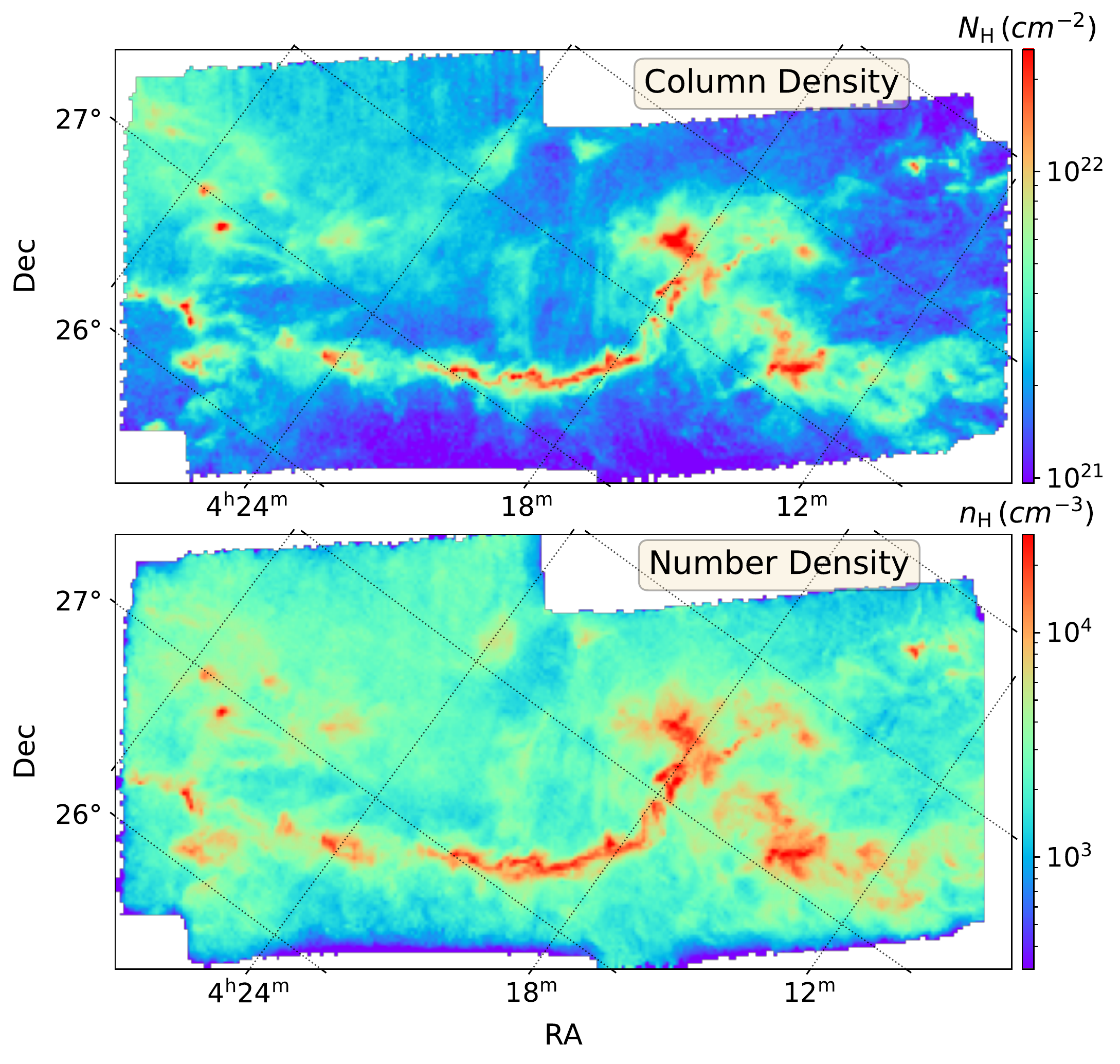}
\caption{$Herschel$ column density map of Taurus B213 ($upper$) and the diffusion model predicted corresponding gas number density ($lower$).}
\label{fig.taurus-B211-test-herschel}
\end{figure*}

In this section, we apply our diffusion model to a $Herschel$-derived column density map of Taurus B213. The $Herschel$ column density map of Taurus B213 is obtained from \citet{2013A&A...550A..38P}, which has a resolution of $18.^{\prime\prime}2$. The column density map is derived based on an optically thin graybody assumption with the function $I_{\nu}=B_{\nu}(T_{d})\kappa_{\nu}\Sigma$, where $I_{\nu}$ is the observed surface brightness at frequency $\nu$, and $\kappa_{\nu}$ is the dust opacity per unit mass. Four data points of spectral energy distributions (SEDs) are included in the fitting, including 160, 250, 350 and 500 \um. 

Taurus B213 is one of the closest star-forming filamentary structures located in the Taurus molecular cloud \citep{2012ApJ...756...12L,2013A&A...550A..38P}. \citet{2013A&A...550A..38P} adopted a power-law radial density profile to fit the column density profile of the B213 filament, and found the central density of the filament to be 
$n_{\rm H,c}=7.5\times 10^{4}$~\cmc. \citet{2012ApJ...756...12L} analyzed multiple transitions of HC$_3$N in B213 filament and constrained the 
number density 
to be $n_{\rm H}\sim 3.6\times 10^{4}$~\cmc. 
We apply our diffusion model to predict the LOS mass-weighted number density and show the predicted map in Figure~\ref{fig.taurus-B211-test-herschel}. The peak number density of B213 filament predicted by the diffusion model is $n_{\rm H,peak}=4.7 \times 10^{4}$~\cmc. 
Note that, the diffusion model predicts the LOS mass-weighted number density but not the actual central peak density of the filament. We adopt the power-law fitting results from \citet{2013A&A...550A..38P} and calculate the LOS mass-weighted number density of the center of B213 filament, which is $n_{\rm H}=4.0\times 10 ^{4}$~\cmc. Consequently, our diffusion model prediction is highly consistent with the estimation from other method. 

{The above example examines the diffusion model's ability to accurately predict number density on real data that were not included in our training set. We further evaluate the diffusion model's performance on a new simulation in Appendix~\ref{Model Evaluation on Different Simulations}, as well as on data with varying physical resolutions in Appendix~\ref{Model Evaluation on Different Physical Scales}. Our findings demonstrate that the diffusion model performs reasonably well in predicting number density on previously unseen data. As a comparison, we also apply \CASItD\ to the Taurus data and report our results in Appendix~\ref{CASItD Performance on Observational Data}. Our findings suggest that the predictions of \CASItD\ exhibit greater blurring, with a smoother density peak at a lower value. This observation implies that \CASItD\ may not be as effective when dealing with unseen data.
}

\subsection{Test on Extinction Maps of IRDCs}
\label{Test on Extinction Maps of IRDCs}

\begin{figure*}[hbt!]
\centering
\includegraphics[width=0.95\linewidth]{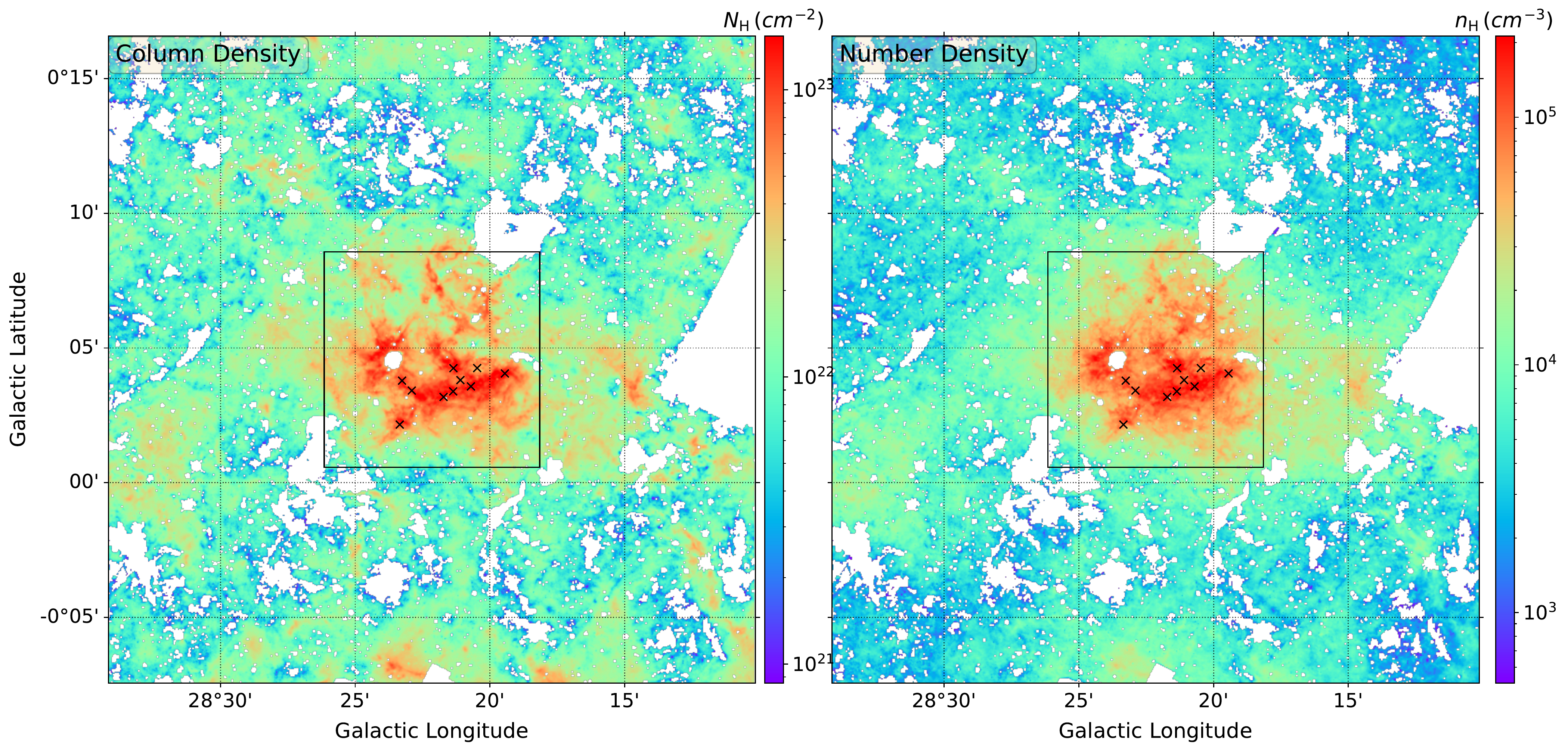}
\caption{Mid-infrared extinction derived column density map of IRDC G28.37+0.07, i.e., Cloud C \citep[$left$, from][]{2013A&A...549A..53K}, and the diffusion model predicted corresponding gas number density ($right$). Cross symbols indicate the locations that have number density estimates by \citet{2022A&A...662A..39E}.}
\label{fig.cloud-c-test-diffusion}
\end{figure*}

\begin{figure}[hbt!]
\centering
\includegraphics[width=0.95\linewidth]{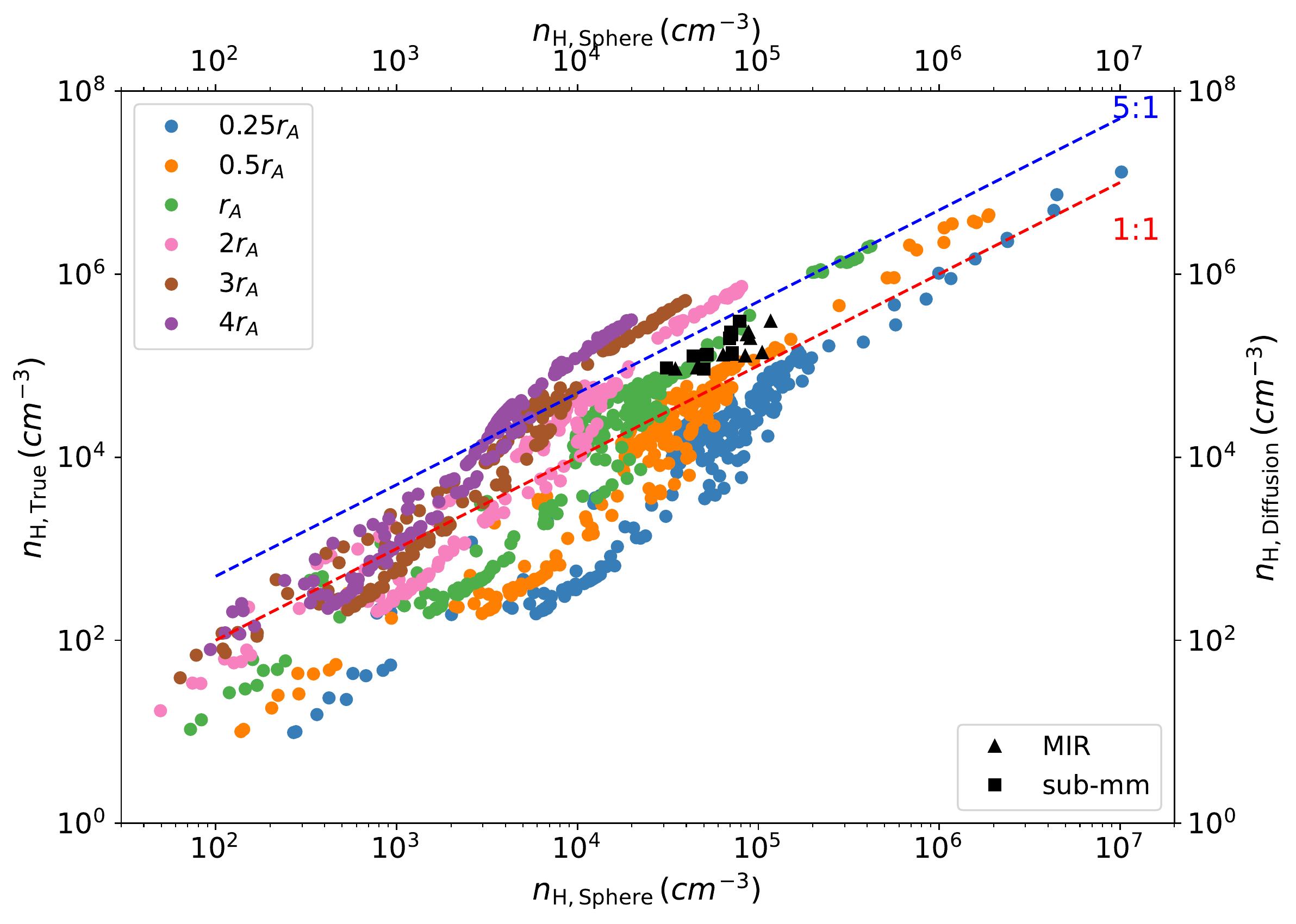}
\caption{Comparison between the true mass-weighted LOS number density and that estimated by assuming a sphere with a certain radius on a simulation sample. Different colors indicate different radii of apertures, where $r_{A}$ is 0.39 pc. Black symbols represent the  comparison between the number density estimates by \citet{2022A&A...662A..39E} and that by the diffusion model in IRDC~G28.37+0.07, i.e., Cloud C. Square and triangle symbols represent different number density estimates in \citet{2022A&A...662A..39E} based on two different data sets, mid-infrared (MIR) extinction from $Spitzer$ and submillimetre (sub-mm) emission from $Herschel$. The red line indicates the 1 to 1 line. The blue line indicates the 5 to 1 line.}
\label{fig.cloud-c-comp-E22}
\end{figure}

\begin{figure*}[hbt!]
\centering
\includegraphics[width=0.95\linewidth]{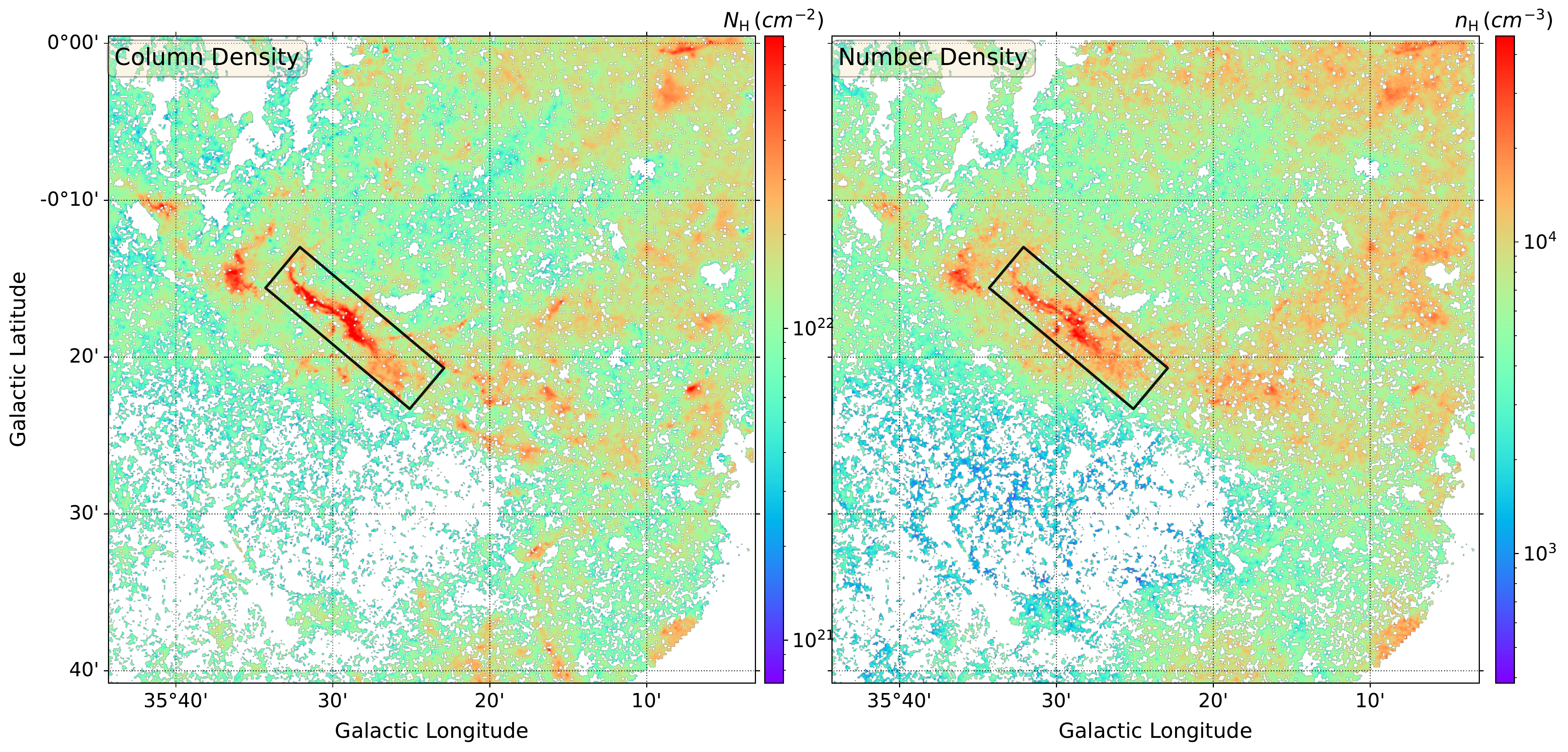}
\caption{Mid-infrared extinction derived column density map of IRDC G35.39-0.33, i.e., Cloud H \citep[$left$, from ][]{2013A&A...549A..53K}, and the diffusion model predicted corresponding gas number density ($right$). The black box indicates the location of the filamentary structure studied in \citet{2014MNRAS.439.1996J}.}
\label{fig.cloud-h-test-diffusion}
\end{figure*}

\begin{figure*}[hbt!]
\centering
\includegraphics[width=0.85\linewidth]{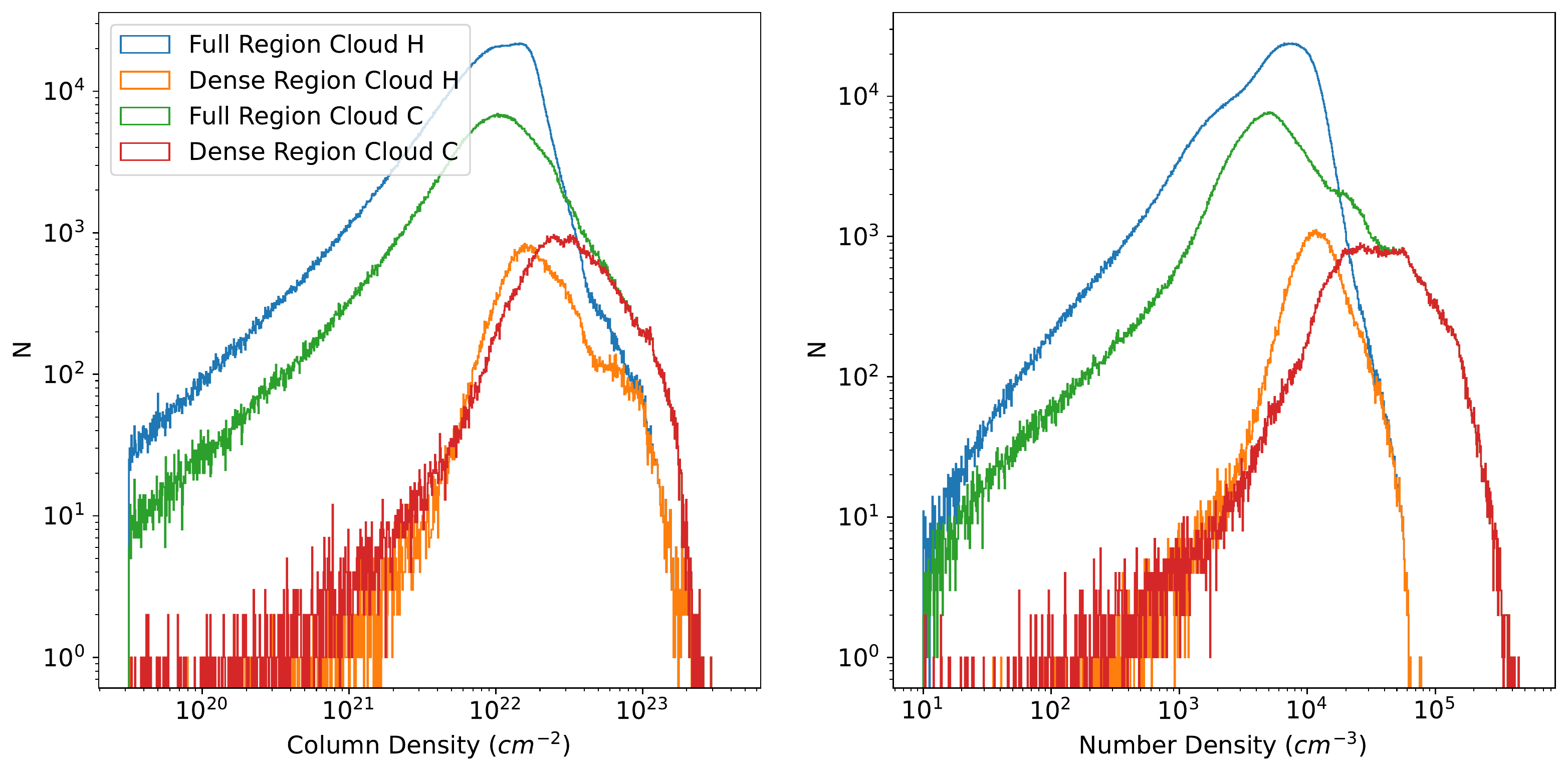}
\caption{Distribution of gas column density and the mass-weighted number density predicted by the diffusion model for both Cloud C and Cloud H. The study of the entire Cloud C and Cloud H regions is represented by the blue and green lines, respectively. The red line corresponds to the study focused on the central dense region of Cloud C as illustrated by the black box in Figure~\ref{fig.cloud-c-test-diffusion}. The orange line indicates the study conducted on the filamentary structure, located within the black box in Fig.~\ref{fig.cloud-h-test-diffusion}.}
\label{fig.cloud_H_pdf_N_n_2panel}
\end{figure*}

In this section, we apply our diffusion model to dust extinction maps of two infrared dark clouds (IRDCs), G28.37+00.07 (Cloud C) and G035.39-00.33 (Cloud H). IRDCs are cold, dense regions that are considered the birthplace of massive stars and star clusters. It is of great significance to study the physical and chemical conditions of IRDCs. Particularly, the number density of IRDCs is one of the most fundamental quantities. 

We adopt the mid-infrared (MIR) extinction (MIREX) map derived column density maps from \citet{2013A&A...549A..53K}, which are derived from $2\arcsec$-resolution 8 \um\ $Spitzer-IRAC$ extinction map \citep{2012ApJ...754....5B}, along with a lower-resolution, larger-scale NIR extinction correction. The column density $N_{\rm H}$ to visual extinction $A_{V}$ relation is $N_{\rm H}=1.9\times 10^{21} {\rm cm^{-2}} (A_{V}/{\rm mag})$. We first apply the diffusion model to the column density map of IRDC G28.37+0.07, i.e., Cloud C, and show the predicted number density in Figure~\ref{fig.cloud-c-test-diffusion}. The number density of Cloud C varies from $10^{3}$ to $10^{6}$ \cmc. \citet{2022A&A...662A..39E} applied spherical geometry with a radius of 0.39~pc to derive the number density from the column density map of Cloud C at ten locations, as indicated in Figure~\ref{fig.cloud-c-test-diffusion}. We compare the number density calculated by \citet{2022A&A...662A..39E} to that by the diffusion model in Figure~\ref{fig.cloud-c-comp-E22}. It is likely that the previous estimation of the number density in Cloud C is systematically underestimated by a factor of {2}, which may have an {slight} impact on the chemical modeling of Cloud C performed by \citet{2022A&A...662A..39E}. To evaluate the impact of the radius of the aperture in spherical geometry approximation, we apply different radii of apertures to calculate the mean number density in each circle and compare that with the true number density calculated from simulations. Figure~\ref{fig.cloud-c-comp-E22} also shows the impact of aperture radius in density estimation. As the aperture goes larger, the mean density calculated with a circle-to-sphere approximation is likely underestimated in dense regions. When the aperture is small, the density calculated with a circle-to-sphere approximation is similar to the true density, which indicates the dense cores likely have a similar scale as this aperture radius. It is worth noting that this scale might only be an upper limit due to the finite resolution of the simulation. At the low-density region, we can see a significant overestimation of density using a circle-to-sphere approximation with a small aperture. This is consistent with intuition, where the diffuse region has a much larger scale on the LOS than 0.39~pc. Consequently, we conclude that the density estimation by with a circle-to-sphere approximation is sensitive to the choice of aperture radius. 

Next, we apply the diffusion model to the column density map of IRDC G35.39-0.33, i.e., Cloud H, which is a highly filamentary cloud. The predicted number density is shown in Figure~\ref{fig.cloud-h-test-diffusion}. The number density of Cloud H varies from $10^{3}$ to $10^{5}$ \cmc. \citet{2014MNRAS.439.1996J} adopted multitransition \13co\ and C$^{18}$O  lines to estimate \h2\ number density of the filamentary structure in Could H. Based on the large velocity gradient (LVG) approximation of the non-local thermal equilibrium (non-LTE) radiative transfer, \citet{2014MNRAS.439.1996J} constrained the 
number density to be $n_{\rm H, avg}=8.3\times 10^{3}-1.2\times 10^{4}$ \cmc. The diffusion model predicts a mean number density of the filamentary structure of $n_{\rm H, avg, diffusion}=1.3\times 10^{4}$ \cmc. The diffusion model predicted mean number density of the entire Cloud H is $n_{\rm H, avg, diffusion}=6.1\times 10^{3}$ \cmc. The predicted number density by the diffusion model is thus consistent with that inferred by an independent non-LTE radiative transfer approach in \citet{2014MNRAS.439.1996J}. 

{It is important to note that our objective is not to reproduce the measurements of individual points obtained by other methods. Rather, our aim is to establish the diffusion model as a useful tool for determining the number density of GMCs over a wide range of values. Such a tool can be useful for various analyses, including statistical analysis of turbulence. An example of comparing the column density distribution with the number density distribution in Cloud C and Cloud H is shown in Figure~\ref{fig.cloud_H_pdf_N_n_2panel}. The dissimilarity between the column density distribution and the number density distribution is evident in the high-density regions. This result may help astronomers in investigating the statistical properties of turbulence more thoroughly in the future. It is noteworthy that the quantity referred to as the number density does not reflect the true density; rather, it represents the mass-weighted number density along the line of sight.
}



\section{Conclusions}
\label{Conclusions}

We have trained the deep learning method diffusion probabilistic models to predict the LOS mass-weighted number density of GMCs from column density maps.  We have tested the diffusion model performance on synthetic test samples and real observational data. Our main findings are as follows:

\begin{enumerate}

\item  The diffusion model is able to predict the LOS mass-weighted number density with higher accuracy (an order of magnitude better) than that from traditional power-law fitting and that from \CASItD.

\item We applied the diffusion model to predict the LOS mass-weighted number density of the Taurus B213 filament from the $Herschel$-derived column density map. The predicted result is consistent with the estimation by the density probe HC$_3$N and by the cylindrical geometry fitting.  

\item We applied the diffusion model to predict the LOS mass-weighted number density of two IRDCs from mid-infrared extinction map derived column density maps. The predicted result is consistent with the estimation by non-LTE radiative transfer modeling. However, application of a local circle-to-sphere conversion appears to underestimate the density by a factor of about {2}, potentially due to the small-scale dense substructure within the volume.

\item {The diffusion model, along with all the other methods of predicting density, encounters difficulties when there are differences between the properties of observational data and those of the training data, such as variations in physical resolutions or dissimilarities in column density distributions. Thus it is important for more precise predictions for the model to be trained with data that most accurately represents the observed system that is being investigated.}


\end{enumerate}

D.X. acknowledges support from the Virginia Initiative on Cosmic Origins (VICO). J.C.T. acknowledges support from NSF grant AST-2009674 and ERC Advanced Grant MSTAR. The authors acknowledge Research Computing at The University of Virginia for providing computational resources and technical support that have contributed to the results reported within this publication. This research has made use of data from the Herschel Gould Belt survey (HGBS) project (http://gouldbelt-herschel.cea.fr). The HGBS is a Herschel Key Programme jointly carried out by SPIRE Specialist Astronomy Group 3 (SAG 3), scientists of several institutes in the PACS Consortium (CEA Saclay, INAF-IFSI Rome and INAF-Arcetri, KU Leuven, MPIA Heidelberg), and scientists of the Herschel Science Center (HSC).

{
\appendix
\section{Model Evaluation on Different Simulations}
\label{Model Evaluation on Different Simulations}

In this section, we apply the power-law fitting result, \CASItD\ and the diffusion model to a new simulation that is conducted with a different code and with different physical conditions. We follow the same simulation setup in \citet{2023ApJ...942...95X}. We conduct ideal MHD simulations with \orion\ \citep{2021JOSS....6.3771L} to model turbulent clouds with periodic boundary conditions and without self-gravity. The simulation box is $5\times5\times5$~pc$^3$.
The magnetic field is initialized along the $z$ direction. The gas is assumed to be an isothermal ideal gas with an initial temperature of 10 K. The three-dimensional Mach number is 10.5, which places the simulated cloud on the line width-size relation, $\sigma _{\rm 1D}=0.72 R^{0.5}_{\rm pc}$ \kms \citep{2007ARA&A..45..565M}. The calculations use a base grid of 256$^3$ without adaptive mesh refinement (AMR).  Simulations are performed with a virial parameter of $\alpha_{\rm vir}=5\sigma_v^2 R/(GM)=2$. We adopt two different mass-to-flux ratios $\mu_{\Phi}=M_{\rm gas}/M_{\rm \Phi}=2\pi G^{1/2}M_{\rm gas}/(BL^2)$, $\mu_{\Phi}$= 1 and 16, which yield Alfv\'en mach numbers of 0.87 and 14, respectively. We show the correlation between the LOS mass-weighted number density and the column density for these turbulent simulations in Figure~\ref{fig.testdatadis_comp}. We also show the correlation between the LOS mass-weighted number density and the column density for all Enzo simulations, i.e., colliding and non-colliding multi-phase GMCs with self-gravity, in Figure~\ref{fig.testdatadis_comp} for comparison. The correlation observed in the \orion\ pure turbulent simulations, which lack self-gravity, differs slightly from the correlation seen in the Enzo simulations that incorporate self-gravity and heating/cooling processes. The range of density variation in the \orion\ simulations is narrower compared to that in the Enzo simulations. However, the overall trend of both correlations is in close accordance.

\begin{figure*}[hbt!]
\centering
\includegraphics[width=0.45\linewidth]{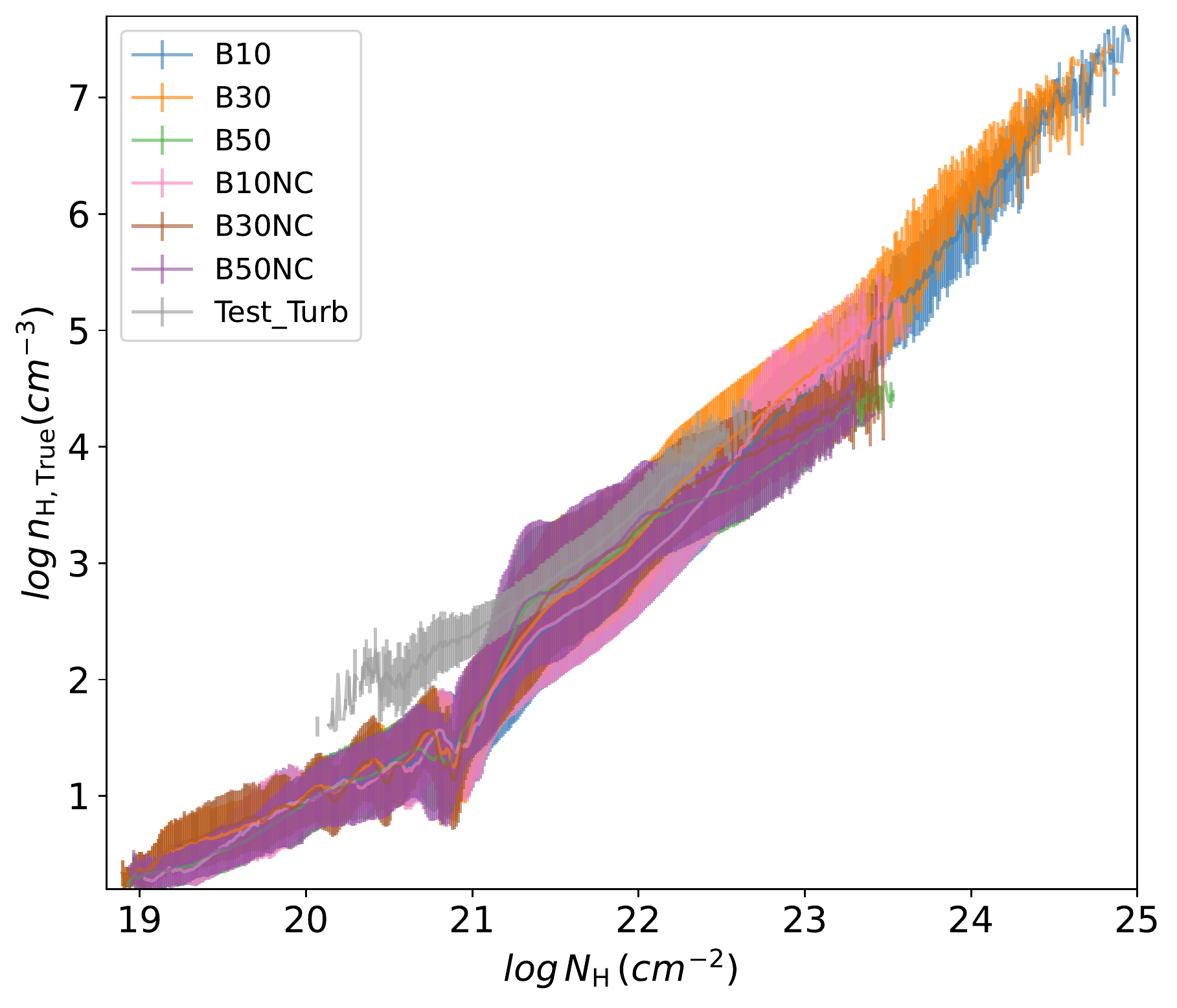}
\caption{Relation between the LOS mass-weighted number density and the column density for different simulations. The lines with error bars represent the mean and their standard deviation of each column density bin.}
\label{fig.testdatadis_comp}
\end{figure*}

\begin{figure*}[hbt!]
\centering
\includegraphics[width=0.95\linewidth]{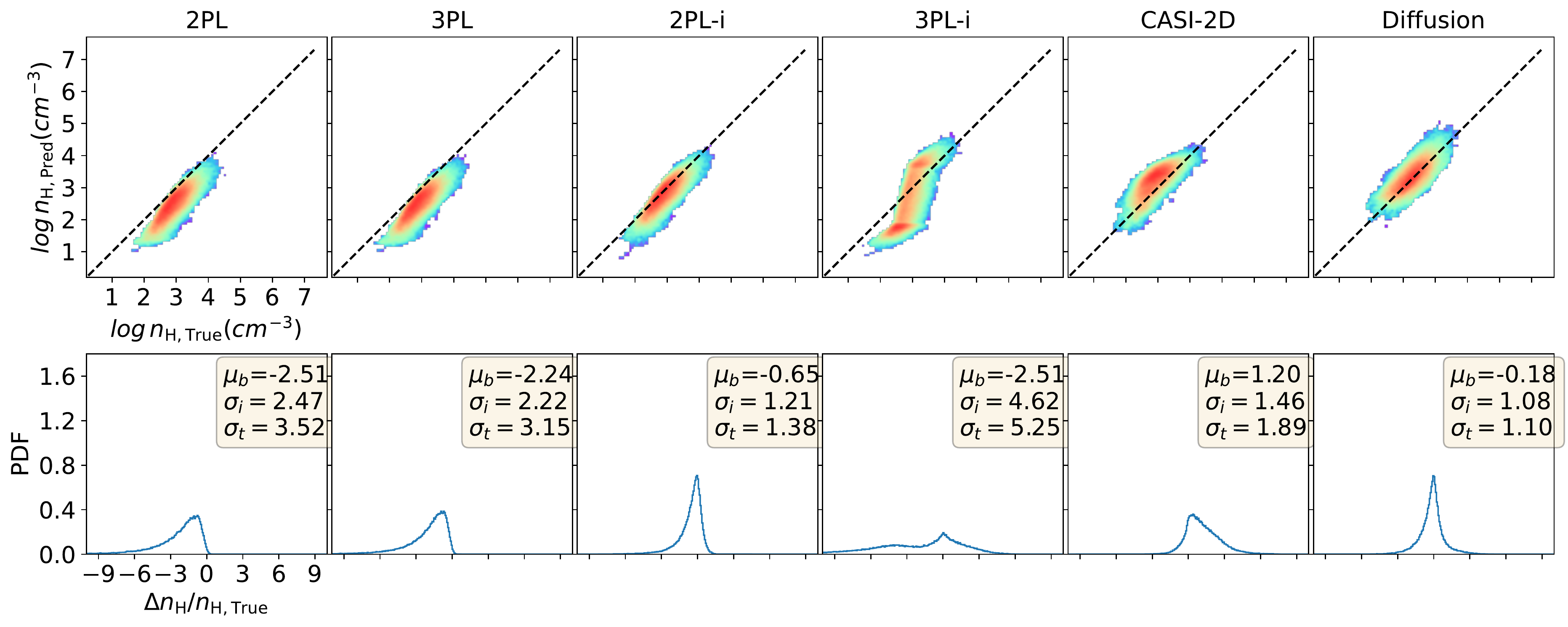}
\caption{Summary of the performance of different approaches to convert the gas column density to number density on \orion\ turbulent simulations.}
\label{fig.diffusion_sum_all_test_turb}
\end{figure*} 

We then apply the same power-law fitting results from \S\ref{Comparison between Different Approaches} and two machine learning approaches to predict the gas number density from their column density. We summarize our results in Figure~\ref{fig.diffusion_sum_all_test_turb}. It is worth noting that all the approaches are derived or trained on the Enzo simulations. This \orion\ turbulent simulation is completely new to these approaches. One can clearly observe that all four power-law conversion approaches are systematically offset from the true value. Similarly, \CASItD\ systematically overestimates the gas number density. On the other hand, the diffusion model is relatively robust in predicting the gas number density with only a small offset. In addition, the error distribution from the diffusion model predictions is more symmetric compared to the other approaches. This highlights the capability of the diffusion model in predicting gas number density on a previously unseen dataset. However, this finding also highlights a potential limitation when applying the diffusion model to observational datasets. Specifically, if the properties of the observational data differ significantly from those of the training data, retraining the model may be necessary to ensure more accurate predictions.

\section{Model Evaluation on Different Physical Scales}
\label{Model Evaluation on Different Physical Scales}

\begin{figure*}[hbt!]
\centering
\includegraphics[width=0.95\linewidth]{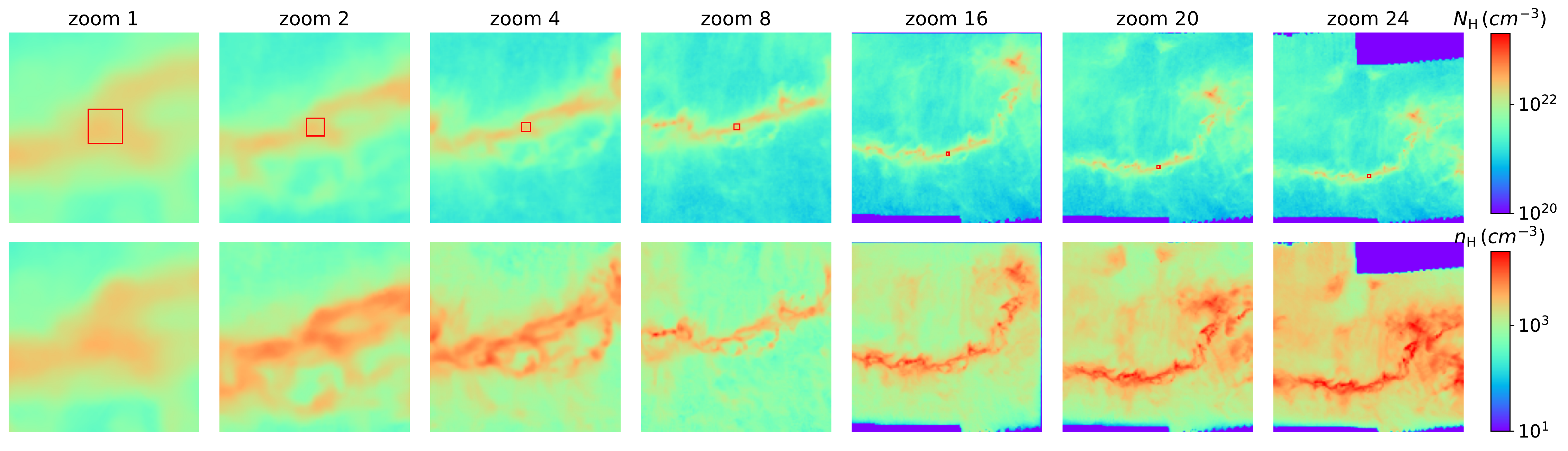}
\caption{Diffusion model performance on different physical scales on a subregion in Taurus. The upper row shows column density maps. The lower row shows the number density maps predicted by the diffusion model. The rectangle indicates the location of the observation by \citet{2012ApJ...745..139L}. }
\label{fig.taurus_cropsub_zoomall_img}
\end{figure*}

\begin{figure*}[hbt!]
\centering
\includegraphics[width=0.45\linewidth]{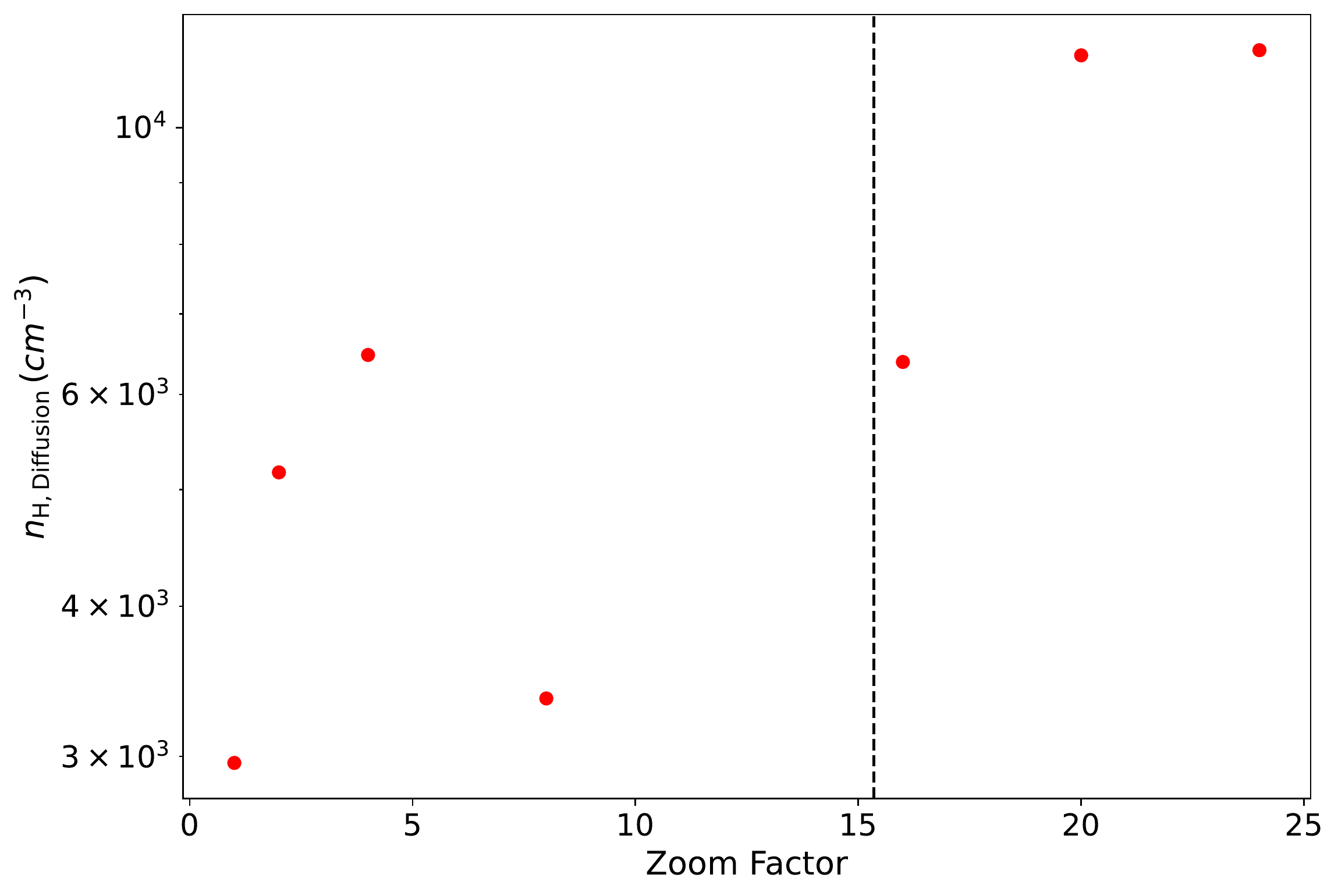}
\caption{Predictions of the number density for the region examined by \citet{2012ApJ...745..139L} at different physical scales. The minimum downsampling factor necessary to match the highest resolution in the training set is indicated by the vertical dashed line. }
\label{fig.taurus_cropsub_zoomall_dotplot}
\end{figure*}

In this section, we evaluate the performance of the diffusion model on observational data at various scales. It is important to point out that the training data covers a particular range of physical resolutions, which spans from 0.315 to 0.25 pc/pixel.
For nearby GMCs, such as Taurus, most column density maps obtained by infrared telescopes, like $Spitzer$ and $Herschel$, have a much higher physical resolution. As a result, to use the diffusion model on such data, we need to downsample it to a lower resolution that is within the range covered by our training set. The column density map of Taurus B213 obtained from \citet{2013A&A...550A..38P} using $Herschel$ has a high physical resolution of 0.002 pc/pixel when adopting a distance of 140 pc for Taurus, as discussed in \S\ref{Test on $Herschel$ Observations of Taurus B213}. However, to use this map with our diffusion model, it needs to be downsampled by a factor of at least 15 to match the physical resolution range covered by our training data. 

Here we assess the performance of the diffusion model on various physical scales using a subregion in Taurus, as illustrated in Figure~\ref{fig.taurus_cropsub_zoomall_img}. The diffusion model likely underestimates the number density by a factor of a few at higher resolutions that are beyond the scope of our training set. To obtain a more accurate assessment, we focus on the same locations observed by \citet{2012ApJ...745..139L} and calculate the average number density in this region across different physical scales. Figure~\ref{fig.taurus_cropsub_zoomall_dotplot} presents the mean density prediction for the same region as \citet{2012ApJ...745..139L} at different physical scales. To achieve the highest resolution within the training set, a minimum downsample factor of 15.4 is required. When the physical resolution is higher than the range of our training set, the diffusion model appears to underestimate the number density by a factor of a few, as compared to the measurements in \citet{2012ApJ...745..139L}. However, when we downsample the column density to a physical resolution of 0.04 and 0.05 pc/pixel, which corresponds to a downsampling factor of 20 and 24 respectively, the model predictions become relatively stable and consistent with the measurements. It is worth noting that we are unable to downsample the Taurus column density map by a larger factor due to the restriction imposed by the diffusion model, which requires an image size of 128$\times$128. 

We proceed to evaluate the performance of the diffusion model on Cloud C and Cloud H at various resolutions. The distances to Cloud C and Cloud H are 5 kpc and 2.9 kpc, respectively, indicating that their physical resolutions are at the upper limit of the resolutions covered by our training set. However, it should be noted that the presence of masked areas in the column density maps of both clouds results in a substantial decrease in the values of downsampled maps when padded with zeros. As a result, we only perform a limited evaluation at a few resolutions as a basic check on the method. 

Figure~\ref{fig.cloud_C_pred_zoom_comp_3panel} and \ref{fig.cloud_H_pred_zoom_comp_4panel} depict the performance of the diffusion model on various physical resolutions of Cloud C and Cloud H. It is evident that the predicted number density reduces by several factors as we downsample the data. To quantify this trend accurately, we take Cloud C as an example, and Figure~\ref{fig.cloud_C_pred_E22compDiff_zoomcomp} displays the diffusion model's prediction on various physical resolutions at locations examined by \citet{2022A&A...662A..39E}. The predicted number density decreases by a factor of two when we downsample the input column density by a factor of two.

\begin{figure*}[hbt!]
\centering
\includegraphics[width=0.99\linewidth]{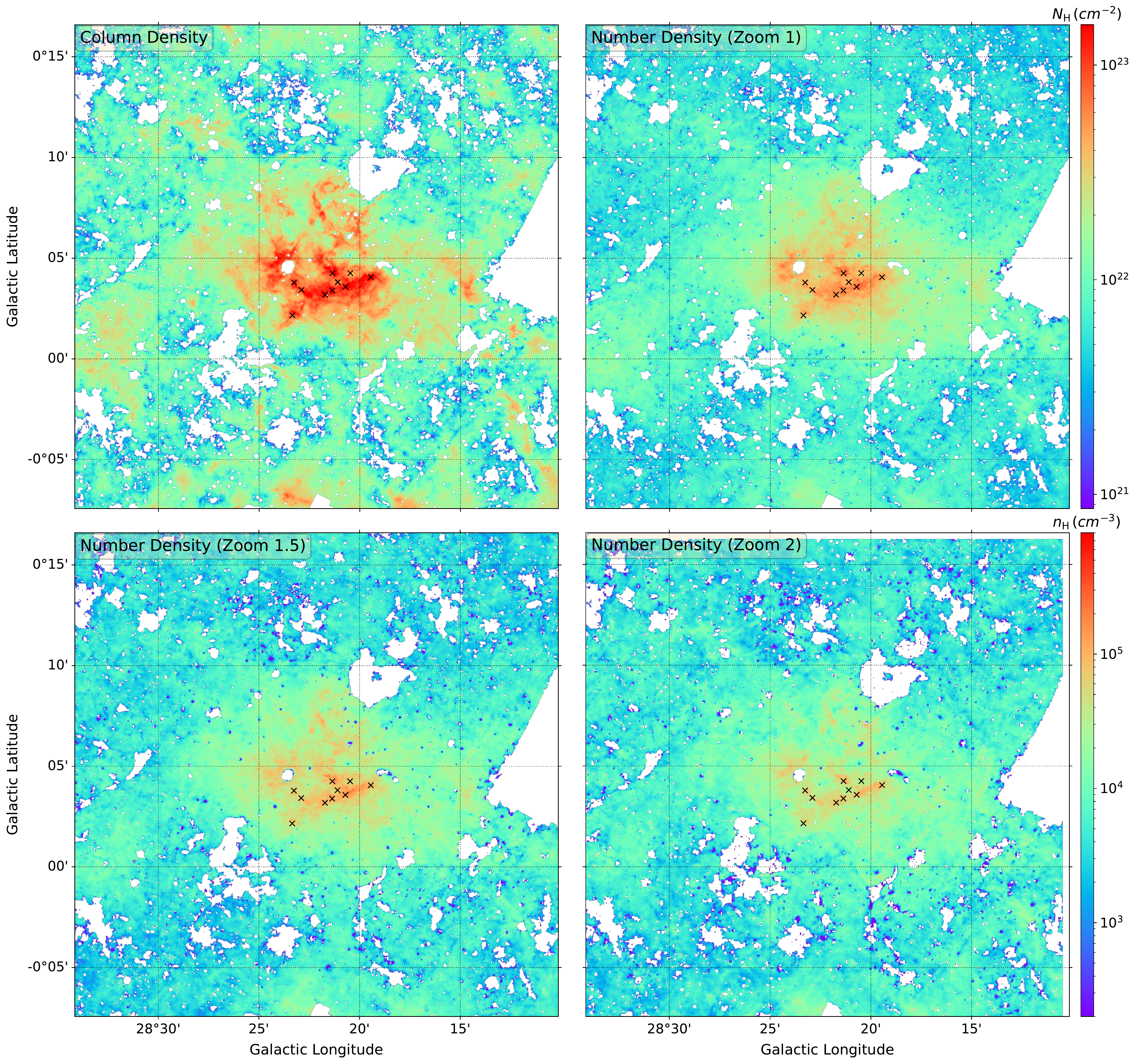}
\caption{Diffusion model performance on different physical resolutions on Cloud C. }
\label{fig.cloud_C_pred_zoom_comp_3panel}
\end{figure*} 

\begin{figure*}[hbt!]
\centering
\includegraphics[width=0.75\linewidth]{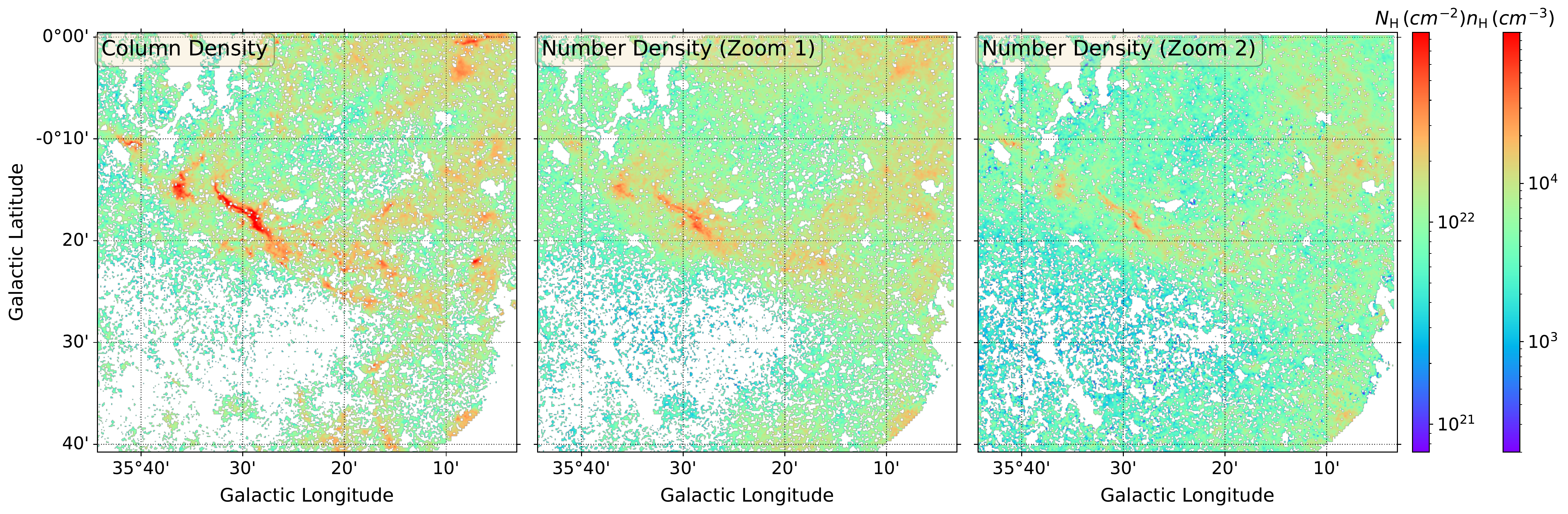}
\caption{Diffusion model performance on different physical resolutions on Cloud H. }
\label{fig.cloud_H_pred_zoom_comp_4panel}
\end{figure*} 

\begin{figure*}[hbt!]
\centering
\includegraphics[width=0.55\linewidth]{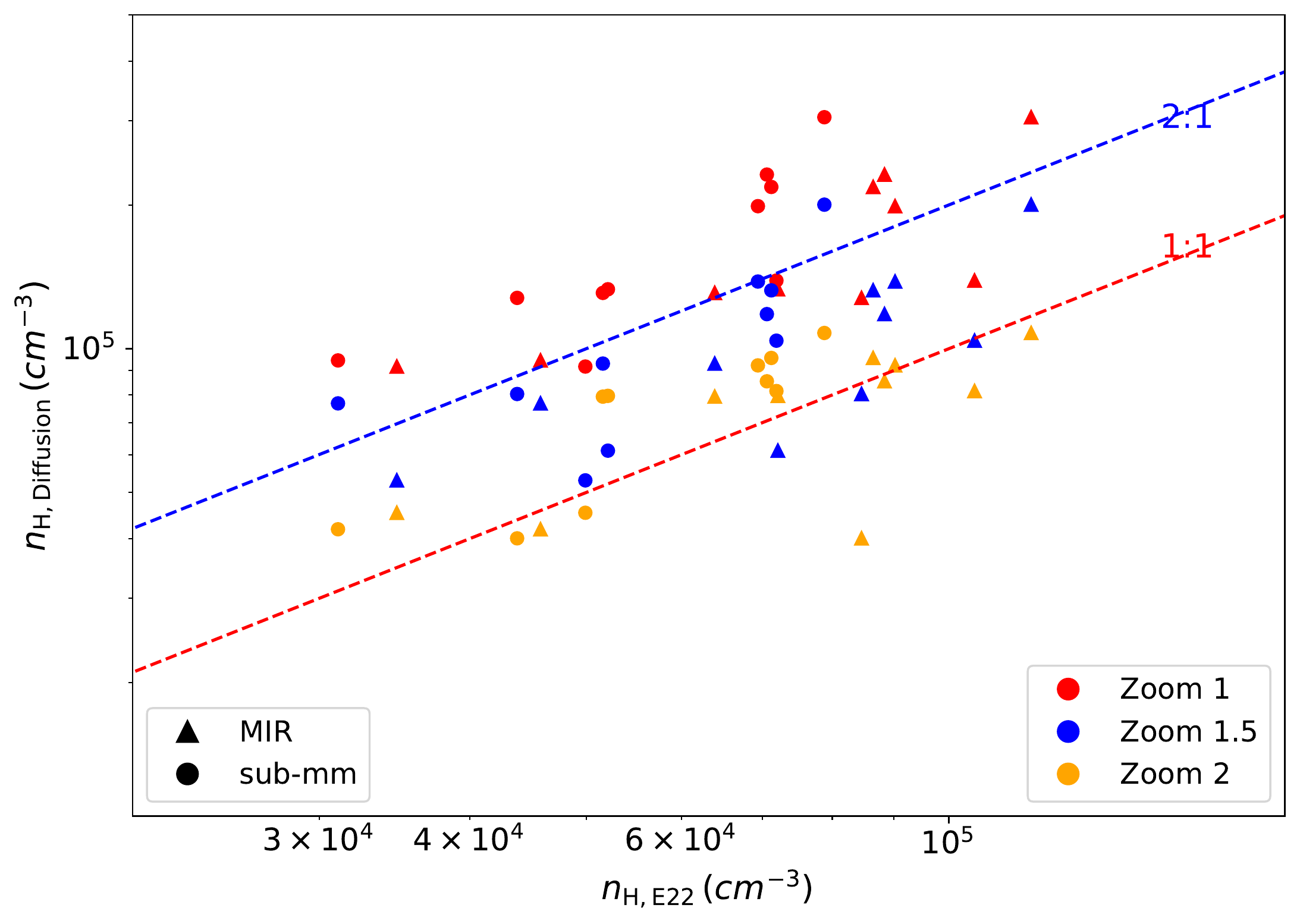}
\caption{Comparison between the number density estimates by \citet{2022A&A...662A..39E} and that by the diffusion model in Cloud C. Square and triangle symbols represent different number density estimates in \citet{2022A&A...662A..39E} based on two different data sets, mid-infrared (MIR) extinction from $Spitzer$ and submillimetre (sub-mm) emission from $Herschel$. The red line indicates the 1 to 1 line. The blue line indicates the 2 to 1 line. Different colors indicate predicted number density at different physical resolutions.
}
\label{fig.cloud_C_pred_E22compDiff_zoomcomp}
\end{figure*}

In light of the differences in physical resolution that may exist between the observational datasets and the training data, it is important to exercise caution when applying the diffusion model to such datasets. In such cases, the diffusion model would need to be retrained with data that is representative of the relevant physical resolutions before being applied to actual observational datasets.

\section{\CASItD\ Performance on Observational Data}
\label{CASItD Performance on Observational Data}

In this section, we provide an overview of the performance of \CASItD\ on observational data. Figures~\ref{fig.taurus_B211_herschel_pred_casi2d} and \ref{fig.cloud_C_H_pred_casi2d} show the number density predictions of \CASItD\ for GMCs. Upon comparison with the predictions of the diffusion model, it is evident that the \CASItD\ predictions are more blurry, with a smoother density peak at a lower value.

\begin{figure*}[hbt!]
\centering
\includegraphics[width=0.75\linewidth]{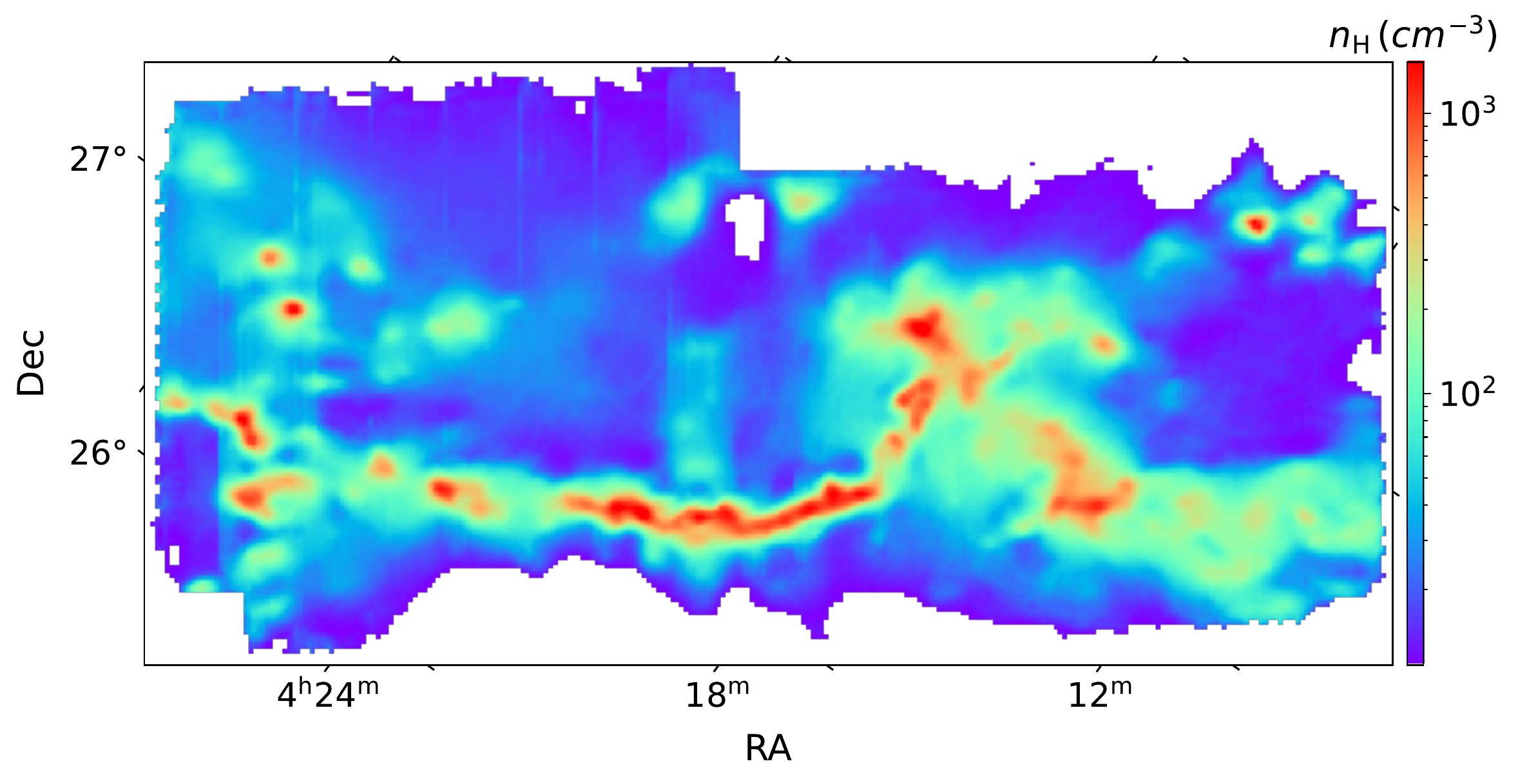}
\caption{Number density predicted by \CASItD\ on Taurus B213. }
\label{fig.taurus_B211_herschel_pred_casi2d}
\end{figure*}

\begin{figure*}[hbt!]
\centering
\includegraphics[width=0.48\linewidth]{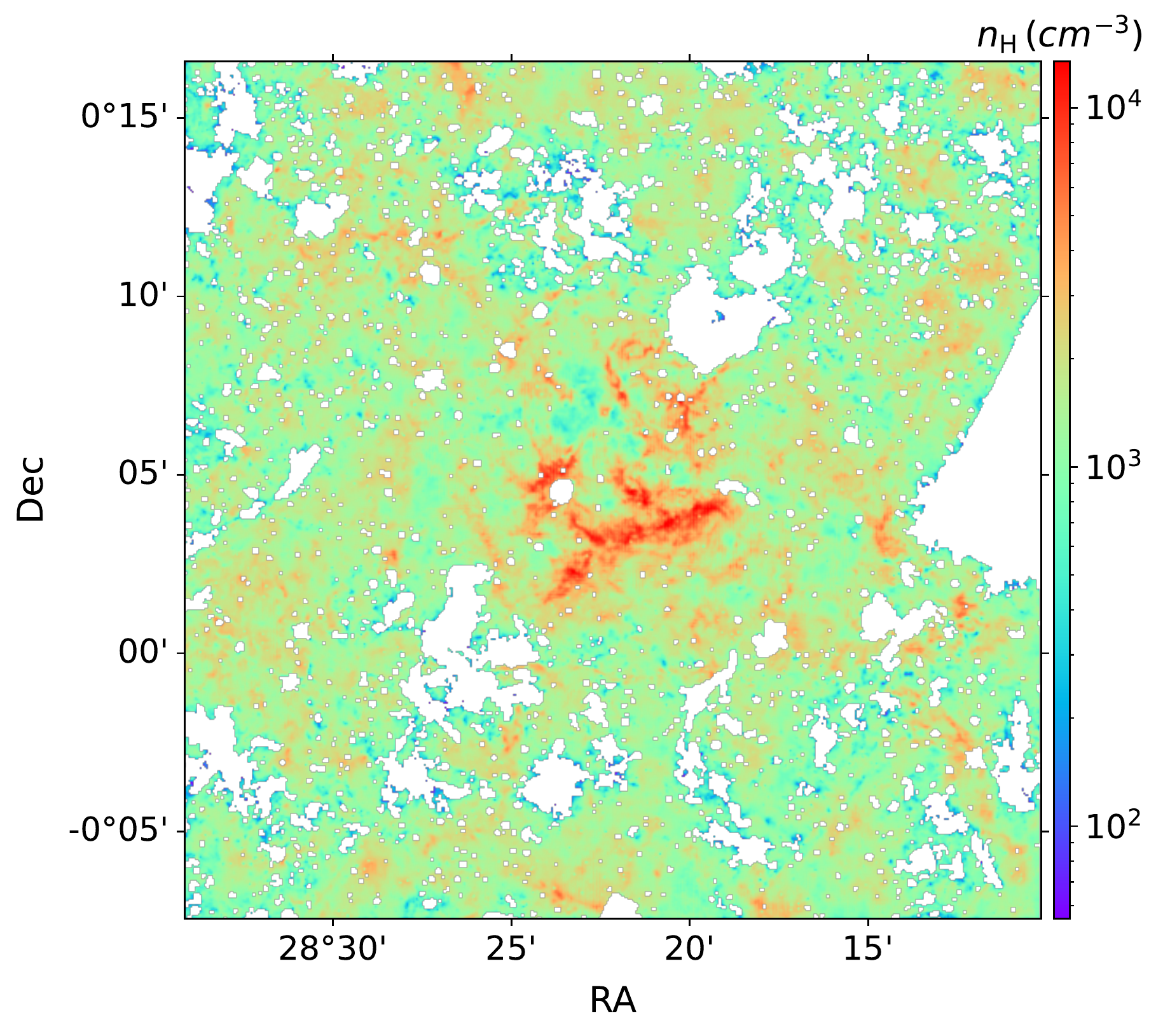}
\includegraphics[width=0.48\linewidth]{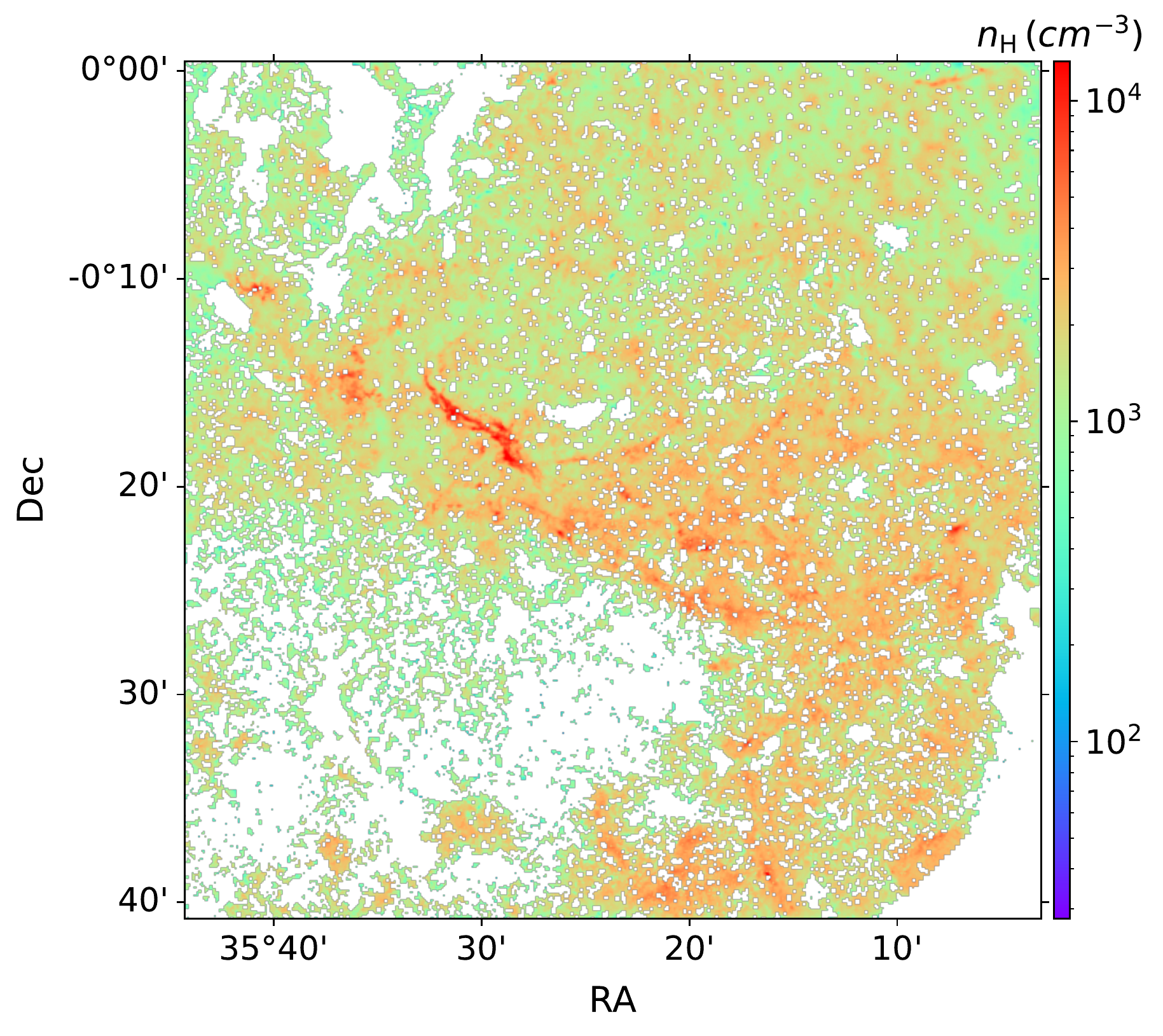}
\caption{Number density predicted by \CASItD\ on Cloud C ($left$) and Cloud H ($right$). }
\label{fig.cloud_C_H_pred_casi2d}
\end{figure*}

}

\bibliographystyle{aasjournal}
\bibliography{references}

\end{CJK*}

\end{document}